\documentclass[letterpaper,twocolumn,10pt]{article}
\usepackage{usenix2019_v3}
\usepackage{filecontents}

\usepackage{amsmath}
\usepackage{amssymb}
\usepackage{enumitem}
\usepackage{graphicx}
\usepackage{subfigure}
\usepackage{amsthm}
\usepackage{hyperref}
\usepackage[ruled,vlined,linesnumbered]{algorithm2e}
\usepackage{bm}
\usepackage{verbatim}

\usepackage[noend]{algpseudocode}
\usepackage{xspace}
\usepackage{url}
\usepackage{breakurl}
\usepackage{color}
\usepackage{soul}
\usepackage{listings}
\usepackage{multirow}
\usepackage{tabularx}
\usepackage{makecell}
\usepackage{booktabs}
\usepackage{tikz}
\usepackage[final]{pdfpages}

\newcommand{\sysname}{Octopus\xspace}
\newcommand{\uromann}[1]{\uppercase\expandafter{\romannumeral#1}}
\newcommand{\romann}[1]{\expandafter{\romannumeral#1}}
\newcommand{\tabincell}[2]{\begin{tabular}{@{}#1@{}}#2\end{tabular}}
\newcommand{\code}[1]{\texttt{#1}\xspace}

\newcommand{\para}[1]{\smallskip\noindent\textbf{#1}}
\newcommand{\type}[1]{$type\ #1$}

\newcommand{\enc}[1]{\text{E}(#1)}
\newcommand{\pr}[1]{\text{Pr}[#1]}
\newcommand{\lap}[1]{\text{Lap}(#1)}

\SetKwInput{KwInput}{Input}
\SetKwInput{KwOutput}{Output}
\SetKwInput{KwStep}{Steps}

\makeatletter
\newcounter{protocol}
\newenvironment{protocol}[1][htb]
  {
   \let\c@algocf\c@protocol
   \SetAlgoCaptionLayout{centerline}
   \begin{algorithm}[#1]%
  }{\end{algorithm}}
\makeatother

\lstdefinestyle{mystyle}{
	keywordstyle=\color{magenta},
	basicstyle=\footnotesize,
	breakatwhitespace=false,
	breaklines=true,
	captionpos=b,
	keepspaces=true,
	numbers=left,
	numbersep=5pt,
	showspaces=false,
	showstringspaces=false,
	showtabs=flase,
	tabsize=2,
	basicstyle=\ttfamily\small
}
\newtheorem{thm}{Theorem}

\newtheorem{defi}{Definition}

\begin{document}

\date{}

\title{\Large \bf \sysname: Privacy-Preserving Collaborative Evaluation of Loan Stacking}

\author{
{\rm Yi Li}\\
Tsinghua University
\and
{\rm Kevin Gao}\\
Tsinghua University
\and
{\rm Yitao Duan}\\
NetEase Youdao
\and
{\rm Wei Xu}\\
Tsinghua University
}

\maketitle

\begin{abstract}
With the rise of online lenders, the loan stacking problem has become a significant issue in the financial industry. One of the key steps in the fight against it is the querying of a borrower's loan history from peer lenders. This is especially important in markets without a trusted credit bureau.
To protect participants' privacy and business interests, we want to hide borrower identities and lenders' data from the loan originator, while simultaneously verifying that the borrower authorizes the query. 
In this paper, we propose \sysname, a distributed system to execute the query while meeting all the above security requirements. 
Theoretically, \sysname is sound. Practically, it integrates multiple optimizations to reduce communication and computation overhead.
Evaluation shows that \sysname can run on 800 geographically distributed servers and can perform a query within about 0.5 seconds on average. 
\end{abstract}

\section{Introduction}
\label{section:introduction}

Within \emph{Internet finance}, peer-to-peer loan services such as Lending Club in the US and CreditEase in China seek to provide accessible and low-cost loans to borrowers, especially to those who would not otherwise be eligible.
These services are very popular in emerging markets, where traditional banks are hesitant to lend to people whose credit cannot be verified by a trusted credit bureau.
For example, in China, there are more than 2000 of loan companies \cite{LENDER_AMOUNT}.  
While they can substitute alternative data like e-commerce purchases and mobile payment transaction logs, they lack the most important data: loan history.
Without such information, malicious borrowers can take out loans from multiple lenders, without the intention of ever paying them back. This is what is commonly referred to as the \emph{loan stacking problem}.

While there are many efforts to build a trusted credit bureau that tracks all credit transactions, like Experian, Equifax, and TransUnion in the US, these firms require the trust of the people.
However, for many markets without such a bureau, or 
with a bureau only providing limited information~\cite{PBCCRC, CREDIT_OVERVIEW}, it is difficult 
to build one because of the trust level people need to impose on it.  In fact, 
recent data breaches like the Equifax breach in 2017~\cite{EQUIFAX_BREACH} raise the question of whether we
should have such bureaus.  

In this paper, we propose \sysname, a cryptography-based solution that reduces the level of trust needed while still meeting the necessary security requirements.
To simplify the discussion, we focus on loan stacking detection: 
An \emph{originator} issues a query to ask as many other lenders as possible about a \emph{borrower}'s outstanding loan amount.  

Cryptographic approaches, such as \cite{PIR, ZKP_HOMO, COMMITMENT, GC, SPDZ}, do not depend on trusted third parties, and reveal no private information other than the pre-negotiated computation results.
However, while applying naive implementations of cryptographic approaches to the loan stacking problem, we face several challenges:
 1) (Authorization)  We need to hide the borrower's identity and at the same time ensure that the query is authorized by the anonymous borrower.
 2) (Scalability in the number of users) Cryptographic operations, especially arithmetical operations on ciphertexts, usually consume substantial computation resources, thus it is challenging to scale to millions of users, the typical size of a lender has. 
 3) (Scalability of the number of lenders) In a large market like China, there are thousands of lenders we want to gather input from. 
 These lenders may be dispersed across the country, and communicate over public networks.
Protocols with high communication cost (e.g. \cite{GC}) or too many interactive rounds (e.g. \cite{SPDZ}) will become impractical.
 4) (Trust on the borrower and originator) Both the borrower and originator have an incentive to lie: the borrower want more loans, while the originator want to get private information as much as possible.  Thus, we need a protocol to catch liars.  

\sysname integrates several security protocols, as well as system design, to provide a solution for the loan stacking problem.  Our large-scale experiments on over 800 servers show that \sysname can perform a privacy-preserving query satisfying the above requirements with about 0.5 seconds on average.
Specifically, the key ideas of \sysname include:
1) We design all the communications between each pair of participants (including the borrower, the originator and lenders) to be \emph{non-interactive}, which means each communication requires at most one round.
2) We adopt \emph{private information retrieval (PIR)}~\cite{PIR, XPIR, OPTIMAL_RATE_PIR} to hide the borrower's identity in the query. We extend the protocol to perform differentially private information retrieval and fully exploit the sparsity of the query keyspace to improve query performance.
3) We design a new protocol that emulates how credit bureaus verify user identities - using secrets in the query result itself. Thus we can authenticate the borrower while hiding the borrower's identity.  
4) We prevent both the borrower and query originator from lying by using \emph{homomorphic commitment scheme}~\cite{ZKP_HOMO, COMMITMENT} - we compute the committed total loan amount collected from the lenders and check the consistency of the sum with the borrower's claim. 
5) We support common query functions, e.g. comparison between the total loan amount and a private threshold of the originator, enabling the borrower to control the amount of information released to the originator. 

We have the following contributions in the paper:
\begin{itemize}[leftmargin=*, itemsep=0mm]
\item We propose \sysname, a system for privacy-preserving loan stacking detection.  \sysname integrates privacy-preserving aggregation, private information retrieval and anonymous authorization in a single system.  Also, with an asynchronous and minimal-round-complexity communication protocol, \sysname achieves high efficiency and scalability.
\item We design a new protocol which performs differentially private information retrieval and utilizes the sparsity to accelerate computation.
\item We design a new protocol to enable anonymous authorization for recursive PIR queries.
\item We implement \sysname and evaluate it in a large-scale, geographically distributed environment with 800 servers. The result shows that \sysname can process requests with a per-request latency of 0.5 seconds on average.
\end{itemize}

\section{Related Work}
To our knowledge, before this work, there has not been practical implementation for privacy-preserving loan stacking detection yet.
On the other hand, there are plenty of works aiming to address similar or related problems.

Many solutions use cryptographic tools (e.g., secure multi-party computation) to build privacy-preserving computation systems. Loan stacking detection can also be easily implemented based on \emph{privacy-preserving aggregation} (i.e., calculating the sum of several private numbers). 
Systems like \cite{SECURE_AGGREGATION, PRIO, P4P, PEM, SEPIA} provide privacy-preserving aggregation for common tasks. For example, Prio~\cite{PRIO} proposes an efficient approach for servers to aggregate data from mobile clients, and provides security against malicious clients using non-interactive zero-knowledge proofs.
However, these systems reveal the final aggregation result directly 
and do not consider the anonymity and identity authentification of the borrower.

In theory, we can use more general privacy-preserving systems, such as \cite{PRIVPY, GLOBAL_MPC, PRIVPY, SHAREMIND, SPDZ} to compute the aggregation. 
However, these works still lack anonymous authorization.  
Also, these systems require synchronous communication, which introduces unnecessary performance costs.
For similar reasons, existing works for privacy-preserving and anonymous database query, such as \cite{PRIVATE_DATABASE}, have super-linear computation complexity, thus not applicable to our case.

While privacy-preserving blockchains are drawing attention~\cite{ZEROCOIN, ZEROCASH, SOLIDUS, ZKLEDGER}, they focus on recording entire transaction histories on a \emph{replicated chain}~\cite{DOUBLE_SPENDING}. We focus on querying \emph{distributed (partitioned)} loan records. 

There are also some solutions based on trusted hardware (e.g., CPU) for privacy-preserving computation, such as \cite{SGX_AGGREGATION, SGX_MPC, HARDWARE_PIR_WEB_SEARCH, HAVEN}. 
However, when we use trusted hardware (such as Intel SGX~\cite{SGX, TSGX}) for privacy-preserving computation, the hardware manufacturers (e.g., Intel) are the roots of trust, and thus should be treated as trusted third parties.
In this paper, we do not assume such trusted third parties, and we focus on software solutions.

\section{System Design}
\label{section:system_design}


\subsection{Problem Formulation}
\label{section:formulation}

\para{Roles. }
There are three roles in the loan stacking problem: a borrower, an originator and $n$ lenders $S_1, S_2, \dots, S_n$. 
A \emph{borrower} may borrow money from one or multiple lenders.
We assume that all the lenders form a consortium, which provides basic services (e.g. user registration and  coordination) but does not touch any private data. 

A user interested in the consortium should register with her real identity. 
An \emph{originator} can inquire a registered user's loan information from the consortium to make some decisions (e.g. whether to lend money to the borrower). 
Note that we allow a user to register to the consortium without becoming a borrower, and we also allow borrowers to not register to the consortium.  
Registration is not binding because users may not want to borrow, but simply display their loan credit or take advantage of other consortium benefits. Registration is not necessary because borrowers don't have to share their loan credits.
This flexibility is essential to bootstrap the system incrementally in an existing loan market.  
In the following sections, we use user and borrower interchangeably.


\para{Geographically-distributed participants. }
\sysname can serve a large loan market, where participants may be dispersed around the world and communicate over the expensive public Internet.
Thus, we need to design the protocol to communicate with minimal rounds to reduce latency, and we want to make all communication asynchronous to tolerate occasional link or server failures.  

\para{Functionality goals.  }
\sysname enables the originator to compute the result of queries of the following form:
\begin{equation}
    f(\sum_{i=1}^{n} x_{ib};t), 
    \label{equation:query}
\end{equation}
where $x_{ib}$'s and $t$ are private inputs to the computation.  In the loan stacking problem, 
$t$ is the private credit limit that the originator assigns to the borrower,
$b$ is the borrower's identity hidden from the lenders, $x_{ib}$ is the loan balance amount at lender $S_i$ for borrower $b$.  The function $f$ is evaluated collaboratively by the borrower and the originator, e.g. a comparison $\sum_i^n x_{ib} < t$.  
A lender $S_i$ may perform some local computation to generate $x_{ib}$ from its raw database. For example, the lender may filter the loan amount and map it to an integer. 
We ignore this pre-process step as the lenders can perform this step locally.
Note that although we present \sysname under the loan stacking application in this paper, Eq.~\ref{equation:query} is generally applicable to other widely used queries~\cite{ZKLEDGER, PEM, SECURE_AGGREGATION}.

\subsection{Architecture Overview}
\label{section:architecture}


\begin{figure}[tb]
	\centering
	\includegraphics[width = 0.35\textwidth]{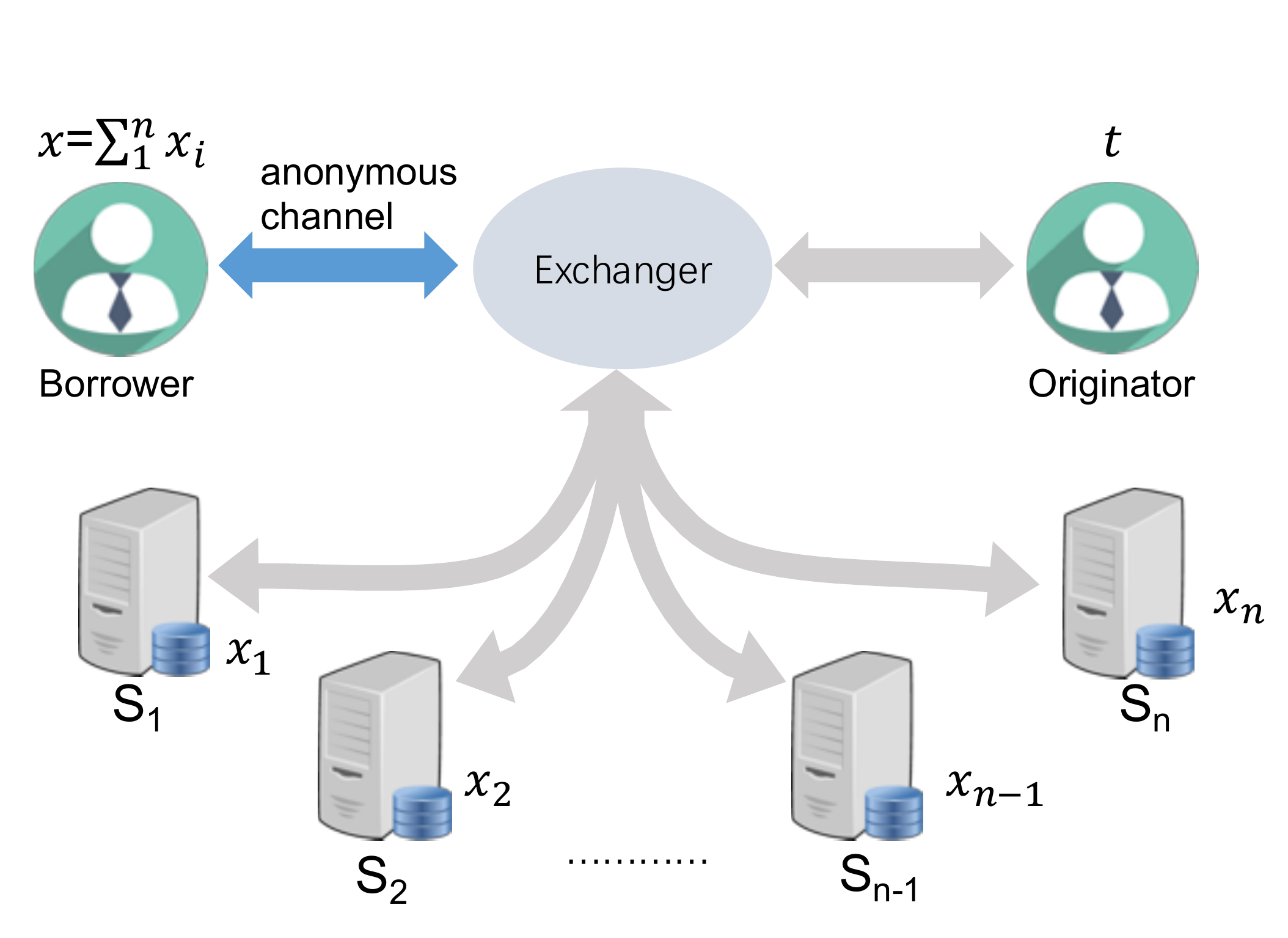}
    \vspace{-0.15in}
	\caption{The architecture of \sysname.}
    \vspace{-0.1in}
	\label{figure:architecture}
\end{figure}

Fig.~\ref{figure:architecture} shows the system architecture of \sysname. 
We introduce an extra role, \emph{exchanger}, as a relay for information.  
It has the following functionalities:
1) The exchanger buffers encrypted messages (but cannot decrypt them), so that participants can communicate asynchronously; 
2) It auxiliarily protects the lenders' databases by adding noise to the responses from each lender;
3) It serves as a central registration database so lenders can synchronize it periodically.
To hide the borrower's identity, we use anonymous channels (e.g. Tor~\cite{TOR}) for communication between the borrower and the exchanger. 

As an optimization, we let the roles pre-share some secrets to enhance efficiency and security. 
For example, when user $u$ registers with the exchanger, she shares a random secret string $\tau_{eu}$ with the exchanger.
When a borrower $b$ borrows money from $S_i$, $S_i$ shares with $b$ not only the loan amount $x_{ib}$ but also a random secret string $\tau_{ib}$.
$\tau_{ib}$ can be used as the seed for generating commitment randomness.
Of course, if a user $u'$ has not borrowed money from $S_i$, we do not assume any shared secret between $u'$ and $S_i$. 

\subsection{Security Goals}
\label{section:security_goal}
At a high-level, the main goal of \sysname is to provide privacy-preserving evaluation of loan stacking.  
On the borrower and originator side, we have the following goals: 

\noindent [\emph{Privacy-B}] Other than the result of Eq.~\ref{equation:query}, the originator learns no information about $x_{ib}$. 

\noindent [\emph{Privacy-O}] The originator keeps the credit limit $t$ private, as each lender evaluates $t$ with her own proprietary algorithm.

\noindent [\emph{Anonymity}] The originator hides the borrower's identity $b$ from other parties, in order to prevent others from competing for the customer.
In this paper, we make the borrower's identity computationally indistinguishable from a group of $N_g$ users. A large enough $N_g$ (e.g., $10,000$) practically conceals the borrower's identity.

In addition, the lenders need two requirements:

\noindent \emph{\romann{1})} [\emph{Privacy-L}] A lender $S_i$ may lend money to many borrowers. Thus, in addition to $x_{ib}$, $S_i$ should keep information about other borrowers private.
More formally, for each borrower $b'$, denoting $I_{ib'}$ as the indicator that $x_{ib'} \ne 0$, we ensure that $I_{ib'}$ is \emph{$(\epsilon, \delta)$-differentially private} to the originator, and protect the exact value of $x_{ib'}$ when $x_{ib'} \ne 0$.

\noindent \emph{\romann{2})} [\emph{Authorization}] As the borrower communicates with the exchanger anonymously and her identity is hidden, an attacker can pretend to be that borrower and collude with the originator to learn about the borrower's loan information (see details in Section~\ref{section:anonymous_authorization}). 
We thus need \emph{anonymous authorization}: the borrower and the originator should collaboratively prove that the borrower who communicates with the exchange anonymously is exactly the one who is queried by the originator.

\subsection{Threat Model}
\label{section:threat_model}
We assume that the exchanger is semi-honest, i.e., the exchanger follows the protocol and does not collude with others, but it is curious about participants' private data.
We also assume that the lenders are semi-honest.
But we assume that the borrower is malicious, i.e., she may lie about her loan history or ideneity.
We also assume that the originator is malicious, as it may send invalid or unauthorized queries to get private information that is not supposed to be revealed to it.
A notable thing is that we assume the borrower and the originator may collude.
We assume an adversary who can monitor all the traffic and control all the connections in the network, in the condition that all the communication is based on secure channels (e.g. SSL). Finally, we assume that the anonymous channels (e.g., Tor) are secure and untraceable.

\sysname is based on standard cryptographic assumptions. We assume secure public-key cryptography systems, homomorphic commitment schemes and  pseudo-random functions. 
Finally, we assume that the originator has a public-secret key pair $(pk_o, sk_o)$ and $pk_o$ is known to all participants.

\section{Solution Overview}
In this section, we first introduce some cryptographic preliminaries for readers unfamiliar with this field, then demonstrate the high-level workflow of \sysname. 

\subsection{Preliminaries}
\label{section:preliminary}

\para{Homomorphic commitment.}
To hide the values of $x_{ib}$'s, we use the \emph{Pedersen commitment} scheme~\cite{COMMITMENT}. 
Given two large primes $p$ and $q$ such that $q | p-1$, we assume that $G_q$ is a subgroup in $\mathbb{Z}_p$ of order $q$. 
Let $g$ and $h$ be two random generators of $G_q$, and define the commitment function as $F(x, r) = g^xh^r \mod p$, where $x \in \mathbb{Z}_q$ is the committed number and $r \in \mathbb{Z}_q$ is a random number.
In this paper, we sometimes abbreviate $F(x,r)$ as $F(x)$.
The commitment function $F$ is \emph{additively homomorphic}: given two commitments $c_1 = F(x_1, r_1)$ and $c_2 = F(x_2, r_2)$, $c_1c_2$ is a commitment that commits to $x_1 + x_2$.
A commitment reveals nothing about the committed value. Moreover, a commitment can be opened in only one way.
In \sysname, each lender $S_i$ commits $x_{ib}$ to $F(x_{ib})$, and the originator cannot open the commitment without $r_i$.

\para{Pseudo-random function.}
A \emph{pseudo-random function (PRF)} is computationally indistinguishable from a truly random function~\cite{CRYPTOGRAPHY_FOUNDATION}.
We denote $\text{PRF}_{s}(x)$ as a PRF function that uses $s$ as the seed and takes $x$ as the input.

\para{Private information retrieval.}
We use \emph{private information retrieval (PIR)}~\cite{PIR} to hide the borrower's identity $b$.
In PIR, there is a \emph{sender} holding a database consisting of $m$ items $\{e_1, e_2, \dots, e_m\}$, and a \emph{receiver} who issues an encrypted query to the database to retrieve an item $e_k$ without revealing the index $k$ to the sender. A \emph{symmetric PIR}~\cite{SYMMETRIC_PIR} further protects the privacy of the sender, i.e., the receiver learner no more information about the database than $e_k$.  



\para{Paillier cryptosystem.}
Paillier is a public key cryptosystem based on the \emph{decisional composite residuosity assumption}~\cite{PAILLIER}.
Given two large primes $p, q$ such that $gcd(pq, (p-1)(q-1)) = 1$, we let $n = pq$ and $g$ be a random integer where $g \in \mathbb{Z}^*_{n^2}$.
To encrypt a message $x \in \mathbb{Z}_n$, we pick a random integer $r \in \mathbb{Z}_n$ and set the ciphertext as $\enc{x} = g^x r^n \mod n^2$.
It is easy to see that Paillier encryption is additively homomorphic: $\enc{x_1}\enc{x_2} = \enc{x_1 + x_2}$.
Another property of Paillier is $\enc{x_1}^{x_2} = \enc{x_1x_2}$.
We use $\boxplus$ and $\boxtimes$ to represent homomorphic addition and homomorphic multiplication respectively. For Paillier, $\enc{x_1} \boxplus \enc{x_2}$ means $\enc{x_1}\enc{x_2}$, while $\enc{x_1} \boxtimes x_2$ means $\enc{x_1}^{x_2}$.

\para{Differential privacy.}
Differential privacy is a rigorous and strong privacy notion~\cite{DIFFERENTIAL_PRIVACY_FOUNDATIONS}. Formally, an algorithm $K$ gives $(\epsilon, \delta)$-differential privacy if for all adjacent datasets $D$, $D^{\prime}$ and all $S \subseteq \text{Range}(K)$, we have 
$\pr{K(D) \in S} \le e^{\epsilon} \cdot \pr{K(D^{\prime}) \in S} + \delta$, 
 where adjacent datasets are two datasets that differ in at most a single record. 
A common method to achieve differential privacy is to add noise following Laplace distribution~\cite{LAPLACE}.
In this paper, we denote a Laplace distribution with mean $\mu$ and variance $2\lambda^2$ as $\lap{\mu, \lambda}$.


\subsection{Solution Overview}
\sysname evaluates the query with three logical processes:

\para{Process 1: Secure aggregation. }
In this process, the originator needs to get the commitment to $\sum_i x_{ib}$ without violating privacy requirements. 
Concretely, the originator uses PIR to collect the commitment to $x_{ib}$ from each lender $S_i$, and checks whether the commitment to $x = \sum_i x_{ib}$ provided by the borrower is consistent with the commitments from the lenders. 

\para{Process 2: Anonymous authorization. }
Since the borrower $b$ and the exchanger communicate anonymously, there needs to be a way for the exchanger to verify the query is authorized by the borrower.
In addition, the originator should prove that the PIR query it sends is valid.
Only after both are verified, can the originator receive the PIR responses from the exchanger and finish the consistency check in Process 1. 

\para{Process 3: Secure evaluation. }
The final goal of the originator is to evaluate a function of the form in Eq.~\ref{equation:query} on the encrypted aggregation (sometimes with its own private input) to get the information helpful for its decision.
We can use ZKP or MPC techniques to enable the borrower and the originator to perform such evaluation. 


We emphasize that we run the three processes in parallel, i.e., process 2 and 3 can start without waiting for process 1 to finish. 
The parallelism reduces the communication rounds and thus computation time.

\section{Process 1: Secure Aggregation}
\label{section:secure_aggregation}

In this process, the originator aggregates information from lenders to evaluate $\sum_i x_{ib}$ in Eq.~\ref{equation:query}, subject to the privacy requirements in Section~\ref{section:security_goal}.



\sysname uses homomorphic commitment scheme to hide the committed data, meeting \emph{[Privacy-B]}. Also, we design a differentially private PIR protocol to ensure \emph{Privacy-L} and \emph{Anonymity}. 
In a high-level view, each lender first generates commitments for its borrowers' data, then the originator uses PIR to retrieve the commitments for the borrower from each lender. 
Then the originator aggregates the commitments and checks the consistency between the aggregated commitment and the one from the borrower to detect possible liars.
As optimizations, we exploit the sparsity of the lenders' databases to accelerate the computation and let the exchanger generate noise commitments for stronger privacy.

\subsection{User Grouping}
We use private information retrieval (PIR)~\cite{PIR} to hide the borrower's identity. 
Theoretically speaking, we can perform PIR on all the registered users. 
However, this is not scalable to hundreds of millions of potential users. 
To address this problem, we divide the registered users into several groups of equal size, and only perform PIR on the corresponding group each time, such that the lenders cannot distinguish a user from others in the same group.
This design is a trade-off between efficiency and privacy. We believe that as long as the group size is large enough, we can achieve adequate privacy.

Specifically, let us assume that there are $N$ registered users, and these users are divided into several groups of size $N_g$.
For a registered user with identity $u$, the exchanger assigns a unique tuple $(gid_u, pid_u)$ to the user, where $gid_u$ is the group id and $pid_u$ is the position of the user in the group. It is obvious that $0 \le gid_u < \lceil N/N_g \rceil$ and $0 \le pid_u < N_g$.
For each $gid$, every lender allocates an array of size $N_g$, and puts a registered borrower $u$ with that $gid$ at position $pid_u$.
A PIR query is performed on the group where the borrower locates, reventing the lenders from distinguishing the current borrowers from the other $N_g-1$ registered users in the same group.

The lenders can synchronize the group ids and user position ids of newly registered users periodically (e.g. daily), and each lender only keeps the information of its own borrowers.
As we have mentioned, it is not necessary for a registered user to have borrowed money, nor is it necessary for a borrower to register with the exchanger.  Thus the synchronization does not lead to any extra privacy issues. 


\subsection{Differentially Private Information Retrieval Utilizing Sparsity (DSPIR)}
\label{section:dspir}

We need a PIR scheme with low communication complexity and practical assumptions.
We use \emph{computationally PIR} (cPIR), which assumes that the participants are limited to probabilistic polynomial-time computations~\cite{CPIR}, and does not need data replication or multiple rounds like \cite{PIR, KEY_PIR}. We choose Paillier as the cryptosystem of cPIR in \sysname.


\para{Naive PIR.  } 
In PIR terminology, we denote \emph{receiver} as the one who sends encrypted queries and receives retrieved data, and \emph{sender} as the one who holds a database and sends responses to receivers.  
In \sysname, the originator is the receiver and every lender is a sender.
Assuming that a sender holds a database with $m$ items $\{e_1, e_2, \dots, e_m\}$ and the receiver wants to retrieve $e_k$, the naive PIR protocol using Paillier cryptosystem works as follows: 
1) The receiver sends a query $q = \{q_1, q_2, \dots, q_m\}$ to the sender, where $q_i$ is a ciphertext of $1$ if $i = k$, or a ciphertext of $0$ otherwise.
2) Upon receiving $q$ from the receiver, the sender calculates $c = \prod_i^m q_i^{e_i}$ and sends $c$ to the receiver.
3) The receiver decrypts $c$ and gets $e_k$.
Step 2 utilizes the homomorphism of the Paillier cryptosystem. 
This simple protocol is easy to implement, but is not practical in our situatioin: since the originator's query size is $O(m)$, the naive protocol has high communication cost.

\begin{figure}[tb]
    \centering
    \subfigure[]{
        \includegraphics[width = 0.15\textwidth]{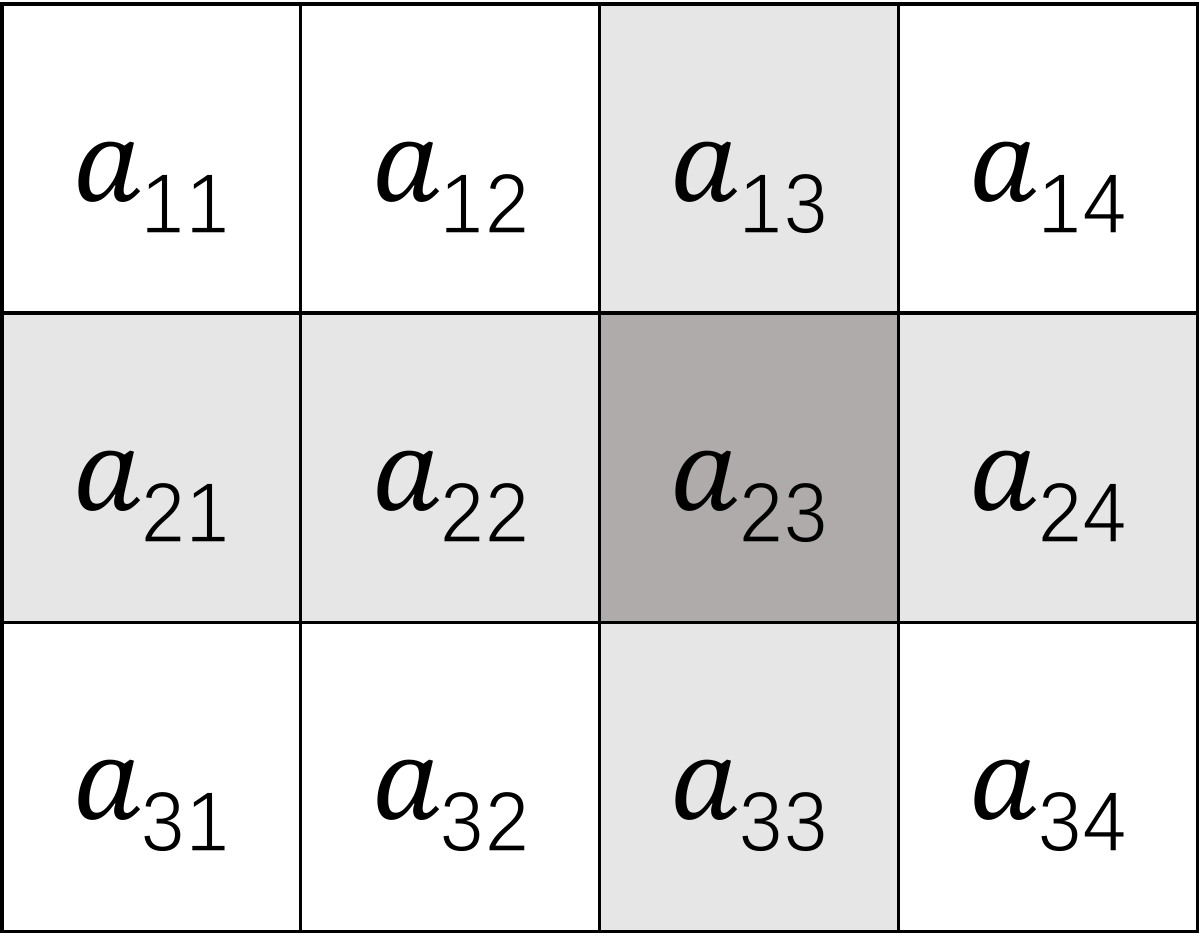}
        \label{figure:pir_full}
    }
    \subfigure[]{
        \includegraphics[width = 0.15\textwidth]{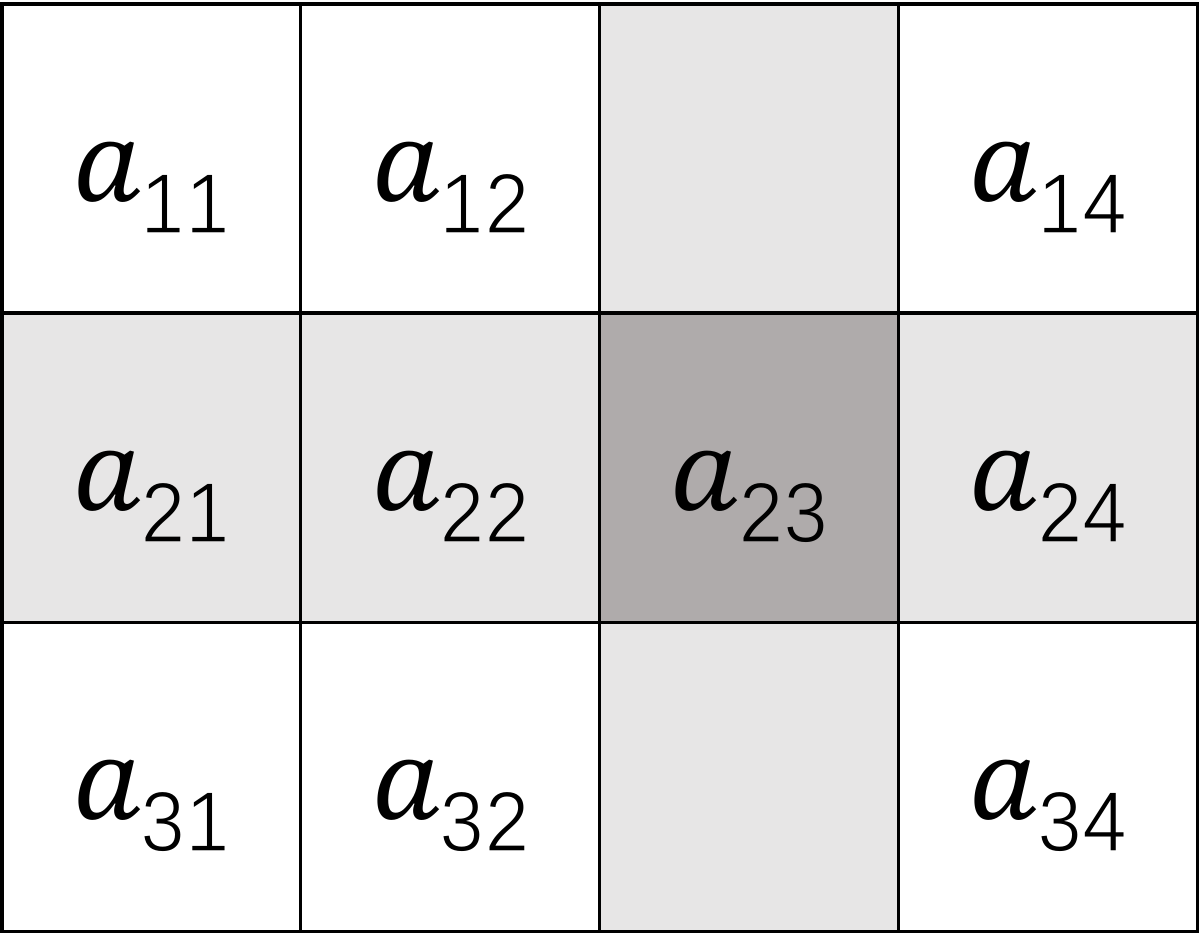}
        \label{figure:pir_sparse_only}
    }
	\vspace{-0.15in}
    \caption{Examples of recursive PIR.}
	\vspace{-0.1in}
    \label{figure:pir_sparse}
\end{figure}

\para{Recursive PIR. } 
To reduce the communication overhead, we can perform PIR recursively~\cite{RECURSIVE_PIR, XPIR}, 
i.e., to query the $k$-th item out of a list of $m$ items, if $m$ can be factorized to $m_1 \times m_2 \times \dots \times m_d$, the receiver can generate a $d$-dimensional query $q = \{\{q_{11}, q_{12}, \dots, q_{1m_1}\}, \{q_{21}, q_{22}, \dots, q_{2m_2}\}, \dots,$ \\ $ \{q_{d1}, q_{d2}, \dots, q_{dm_d}\}\}$, where $q_{ij}$ is a ciphertext of $1$ if the $k$-th item falls into the position $j$ of the $i$-th dimension, or a ciphertext of $0$ otherwise.
Then the receiver only needs to send $\sum_{i=1}^{d}m_i$ ciphertexts to the sender, while the receiver should send $m = \prod_{i=1}^d m_i$ ciphertexts in a non-recursion version. 
Fig.~\ref{figure:pir_full} shows an example. To retrieve $a_{23}$, the receiver sends $3 + 4 = 7$ ciphertexts. The sender first retrieves the 2nd row, then retrieves the 3rd column from the retrieved row.

The \emph{expansion factor}, or the ratio of the size of ciphertext to the size of plaintext, also affects the communication cost.  
For example, 1024-bit Paillier encrypts a 1024-bit plaintext to a 2048-bit ciphertext, so the expansion factor is $f = 2$.
It can be seen that for a $d$-dimensional recursive PIR query, the expansion factor is $f^d$.
Therefore, for $l$-bit Paillier, which means the bit-length of the plaintext is $l$, the total communication cost of a $d$-dimensional recursive PIR query is $(\sum_{i=1}^{d} m_i + f^d)l$. Thus we can choose proper cryptosystems (e.g. Paillier) and parameters (e.g. $d$ and $m_i$'s) to minimize communication cost.


\para{Sparsity-aware PIR - first version.  }
As each lender $S_i$ only stores loan information of its own borrowers, its array of the borrowers' information is likely to be very sparse for all groups.
For example, if only $1/10$ of people have borrowed money from lenders, the \emph{sparsity}, or the ratio of the empty items, of most lenders will be over $0.9$.
\sysname utilizes the sparsity to accelerate the PIR with a simple idea: just skip the empty items when generating responses. 

Specifically, for a sparse array $A$ with $m$ items, assume that $m'$ of the items are non-empty (in our situation, this means $m = N_g$ and only $m'$ users of a group have borrowed money from that lender). We represent $A$ as $\{ind_1:a_1, ind_2:a_2, \dots, ind_{m'}:a_{m'}\}$, where each $(ind_j, a_j) (1 \le j \le m')$ is an index-value pair. In this paper, each $ind_j$ corresponds to a $pid$, while each $a_j$ corresponds to a commitment to some $x_{ib}$.
Algorithm~\ref{algorithm:pir_sparse_recursive} summarizes the sparsity-aware PIR algorithm run at the lenders. 
Intuitively, for an array $A^*$ of size $m^*$, if we want to retrieve the item at position $ind$, we can first aggregate every $row\_len$ items into a group and treat the array as a $row\_len \times (m^*/row\_len)$ matrix (line 4). Next, we calculate the coordinate of $A^*[ind]$ and denote it as $(r, c)$ (line 7-8), then retrieve the $r$-th row. Next, we apply this process recursively to retrieve the $c$-th item of the retrieved row (line 3 to line 17). 
We skip the empty items to reduce computation overhead (line 6-12).
Finally, the algorithm returns the retrieved item. 
The computation complexity is proportional to the number of non-empty items (i.e., the number of registered users who have borrowed money from the lender).
Although the idea is straightforward, there remain two issues:

First, as the sender skips empty items, the receiver can deduce extra information about the array by comparing the result ciphertext to certain numbers. For example, Fig.~\ref{figure:pir_full} shows a $3 \times 4$ array, of which all items are non-empty, while in Fig.~\ref{figure:pir_sparse_only}, the 3rd column only contains the inquired item $a_{23}$.
For Fig.~\ref{figure:pir_full}, when the sender retrieves the 2nd row, as $q_{ij} = g^{x_{ij}}r_{ij}^n$ where $x_{ij} = 0 \text{ or } 1$, the sender gets $\enc{a_{23}} = g^{a_{23}}(r_{11}^{a_{13}} r_{12}^{a_{23}} r_{13}^{a_{33}})^n$.
However, for Fig.~\ref{figure:pir_sparse_only}, when the sender retrieves the 2nd row, the sender gets $\enc{a_{23}} = g^{a_{23}}(r_{12}^{a_{23}})^n$.
Thus, after receiving $\enc{\enc{a_{23}}}$, the receiver first decrypts the outermost encryption and gets $\enc{a_{23}}$. By comparing $g^{a_{23}}(r_{12}^{a_{23}})^n$ and $\enc{a_{23}}$, the receiver can deduce whether the 3rd column contains other non-empty items or not.

\begin{figure}[tb]
    \centering
    \subfigure[]{
        \includegraphics[width = 0.14\textwidth]{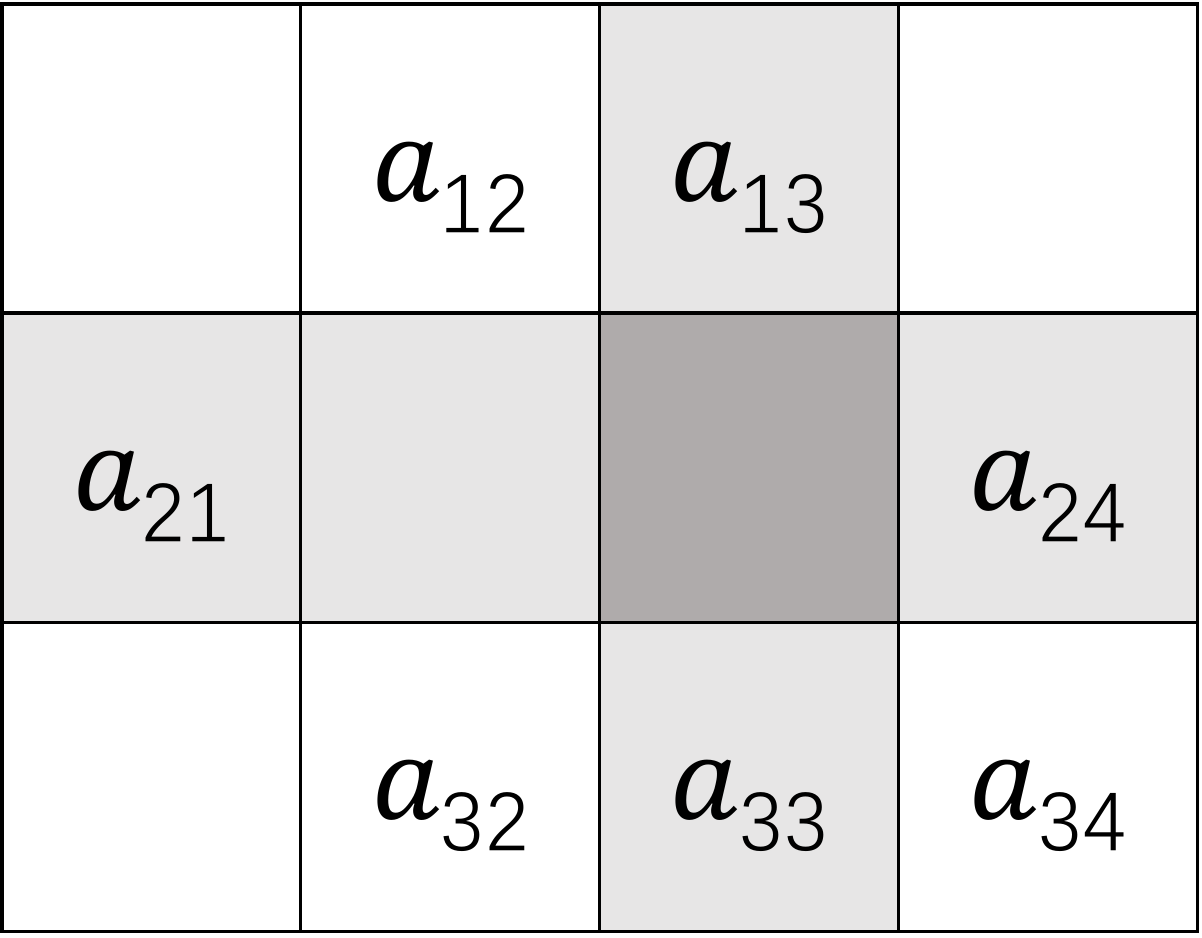}
        \label{figure:pir_sparse_zero1}
    }
    \subfigure[]{
        \includegraphics[width = 0.14\textwidth]{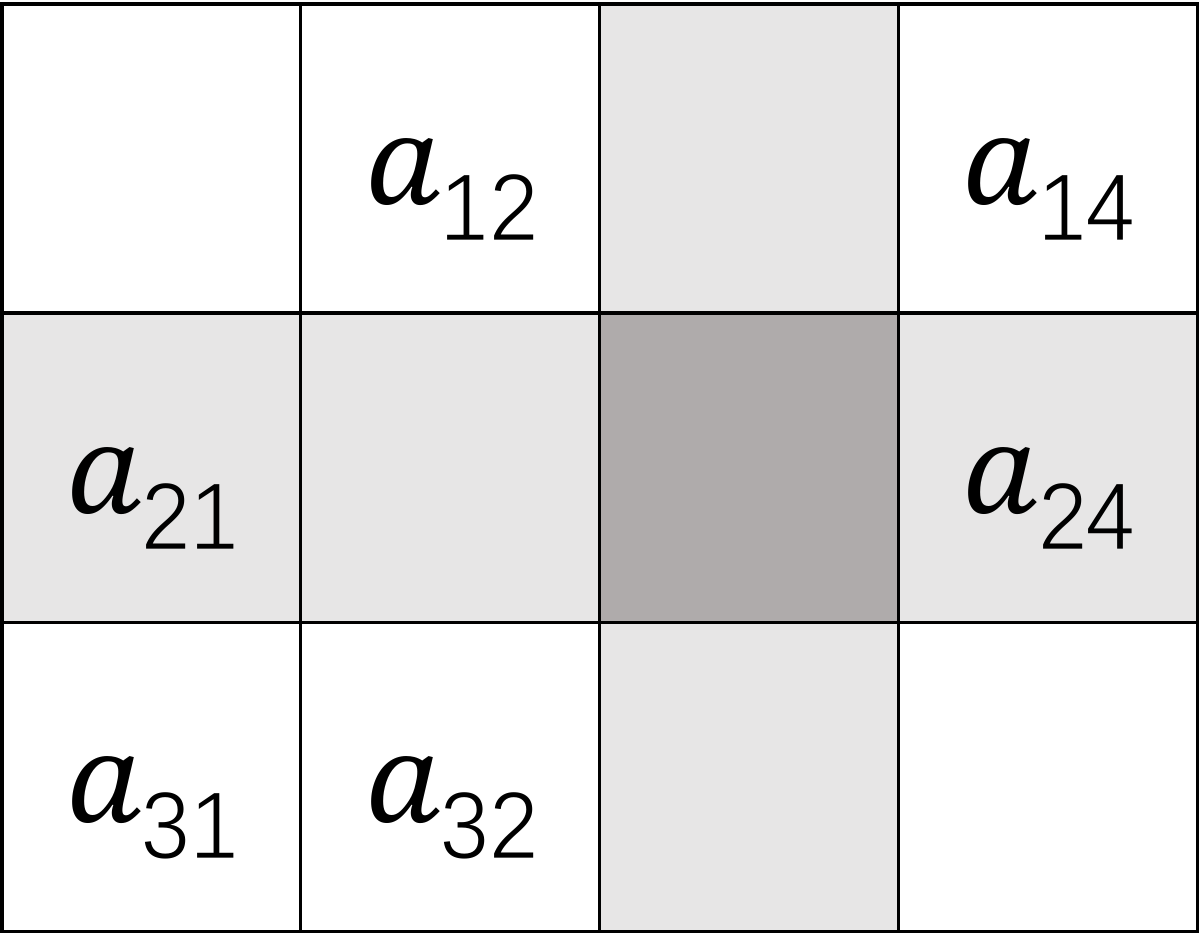}
        \label{figure:pir_sparse_zero2}
    }
    \subfigure[]{
        \includegraphics[width = 0.14\textwidth]{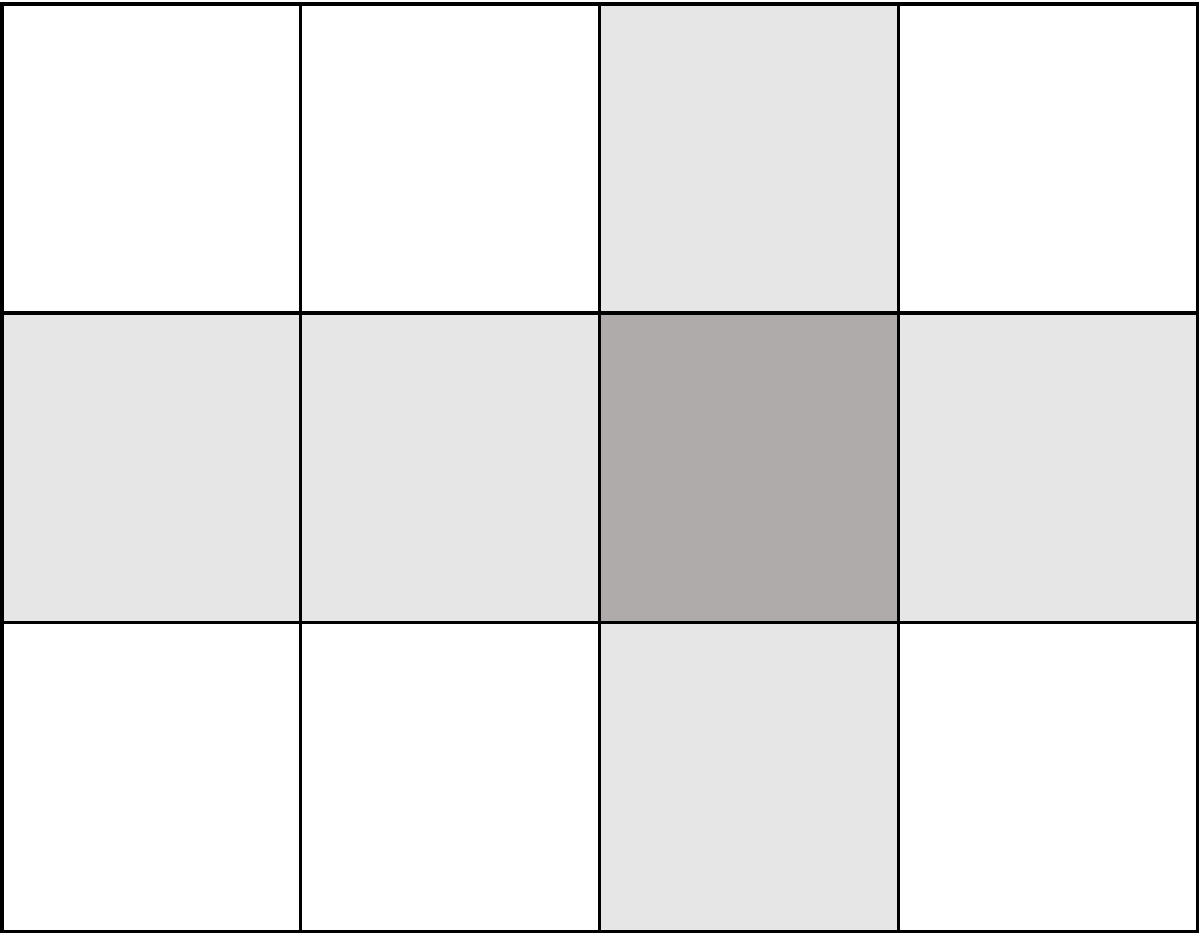}
        \label{figure:pir_sparse_zero3}
    }
	\vspace{-0.15in}
    \caption{Examples of PIR utilizing sparsity.}
	\vspace{-0.1in}
    \label{figure:pir_sparse_zero}
\end{figure}

Second, the receiver can deduce extra information by checking whether the decryption returns 0.
For example, a query for Fiq.~\ref{figure:pir_sparse_zero1} returns $\enc{\enc{0}}$. This is because $a_{23}$ is empty and thus skipped.
Similarly, a query for Fig.~\ref{figure:pir_sparse_zero2} returns $\enc{0_1}$, where $0_1$ is a $0$ string whose the length equals to the length of a ciphertext (e.g. 2048 bits for Paillier with a 1024-bit private key).
Fig.~\ref{figure:pir_sparse_zero3} shows an entirely empty array, and we set 0 as the output (line 19 in Algorithm~\ref{algorithm:pir_sparse_recursive}).
Generally, with a $d$-dimensional query for retrieving an item $e$, there are $d+2$ possibilities for the result of Algorithm~\ref{algorithm:pir_sparse_recursive}: $0$, $\enc{0_{d-1}}$, $\text{E}^2(0_{d-2})$, $\dots$, $\text{E}^{d-i}(0_{i})$, $\dots$, $\text{E}^{d-1}(0_{1})$, $\text{E}^d(0)$, $\text{E}^d(e)$, where $\text{E}^i(\cdot) = \enc{\text{E}^{i-1}(\cdot)}$ and $0_{i}$ is a $0$ string of the same length as $\enc{0_{i-1}}$. Although these ciphertexts are of equal length, the receiver can decrypt them and discover their types.
Thus, different kinds of arrays result in different kinds of outputs, revealing extra information.

\begin{algorithm}[tb]
\footnotesize
\DontPrintSemicolon
    \KwIn{The query $q = \{\{q_{11}, q_{12}, \dots, q_{1m_1}\}, \{q_{21}, q_{22}, $ \\ $ \dots, q_{2m_2}\}, \dots, \{q_{d1}, q_{d2}, \dots, q_{dm_d}\}\}$, 
    the query dimension $d$, and the sparse array/dataset $A = \{ind_1:a_1, ind_2:a_2, \dots, ind_{m'}:a_{m'}\}$.}
    \KwOut{The cipertext of the queried item.}
    Let $A^* = \{ind_1^*:a_1^*, ind_2^*:a_2^*, \dots, ind_{m'^*}^*:a_{m'^*}^*\}$ be a copy of $A$. \\
    Initialize $m^*$ as the capacity of $A^*$ (i.e. $m^* := m$) and $m'^*$ as the number of non-empty items in $A^*$ (i.e. $m'^* := m'$). \\
    \For{$i=1; i \le d; i = i + 1$}
    {
        $row\_len = m^* / m_i$ \\
        $tmp\_A^* = \text{empty\_set}()$ \\
        \For{$j=1; j \le m'^*; j = j + 1$}
        {
            $r = \lfloor ind_j^* / row\_len \rfloor$ \\
            $c = ind_j^* \mod row\_len$ \\
            \If{$tmp\_A^*[c]$ \text{is empty}}
            {
                $tmp\_A^*.\text{append}(c : q_{ir} \boxtimes {a_j^*})$ \\
            } \Else {
                $tmp\_A^*[c] = tmp\_A^*[c] \boxplus (q_{ir} \boxtimes {a_j^*})$ \\
            }
        }
        \For{$(ind, a) \in tmp\_A^*$}
        {
            $tmp\_A^*[ind] = tmp\_A^*[ind] \boxplus \enc{0}$ \\
        }
        Assign $A^* = tmp\_A^*$ \\
        $m^* = m^* / m_i$ \\
        $m'^* = tmp\_A^*.\text{non\_empty\_count}()$
    }
    \If{$A^*$ is empty}{\textbf{return} $0$}
    \Else {\textbf{return} the item in $A^*$}
\caption{Sparsity-aware PIR algorithm.}
\label{algorithm:pir_sparse_recursive}
\end{algorithm}

We address the first issue by introducing extra randomness to the result. 
In line $16$, we add a random ciphertext of $0$, i.e. $r^{n}$, to mask each ciphertext.
Thus, under the \emph{decisional composite residuosity assumption}~\cite{PAILLIER}, the receiver cannot infer extra information from the ciphertexts without knowing the values of the introduced randomness.

\para{Enhancing security with noise responses.  }
Our solution to the second issue is more involved.  We let the exchanger introduce perturbations: the exchanger adds noise responses to the lenders' responses, to make the originator unable to distinguish the response type of a lender.
Apart from $0$, the other possible outputs of Algorithm~\ref{algorithm:pir_sparse_recursive} are of equal length. To prevent the exchanger from learning additional information from the outputs, the lenders replace the output $0$ with $\text{E}(0_{d-1})$.
For $i = 1, 2, \dots, d$, we say that a response is of \type{i} if it is of the form $\text{E}^{d-i}(0_i)$. And we say a response is of \type{0} if it is a ciphertext of a commitment, i.e.  $\text{E}^d(F(x))$.
Then the exchanger generates noise responses of these types. The noise responses of \type{1} to \type{d} can be generated directly by the exchanger using the public key of the originator. 
To hide the number of responses of \type{0}, the exchanger generates commitments to $0$, namely $F(0)$, and then encrypts each of them as $\text{E}^d(F(0))$.
As long as the originator cannot distinguish commitments to $0$ from other commitments, it cannot tell whether a commitment is from the exchanger or the lenders. 
Meanwhile, as the noise responses are ciphertexts of $0$ strings or commitments to $0$, they do not affect the final computation result.

Specifically, the exchanger adds \emph{Laplace noise} to achieve \emph{differential privacy}~\cite{DIFFERENTIAL_PRIVACY_FOUNDATIONS}.
The type count the originator gets can be represented as a vector $\vec{v} = (n_0, n_1, \dots, n_{\hat{d}})$, where $\hat{d}$ is the number of possible response types ($\hat{d}=d$ in the simple case) and $n_i$ is the number of responses of \type{i} ($0 \le i \le \hat{d}$).
We first consider a simple case where a newcoming originator sends a query without any prior knowledge of the dataset.
The following theorem states how much noise is required in this case (see Appendix~\ref{appendix:proof_dp} for the proof):

\begin{thm}
\label{thm:differential_privacy}
If the exchanger generates $\widetilde{n_i}$ noise responses of \type{i}($0 \le i \le \hat{d}$), where $\widetilde{n_i} \sim \lceil \max(0, \lap{\mu, \lambda}) \rceil$ for each $i$, then the exchanger makes the type of the response from each lender $(\epsilon, \delta)$-differentially private to the originator, where $\epsilon = \frac{2}{\lambda}$ and $\delta = e^{\frac{1-\mu}{\lambda}}(1- \frac{1}{4}e^{\frac{1-\mu}{\lambda}})$.
\end{thm}

In general cases, however, the type count vector $\vec{v}$ is not enough, due to two reasons:
1) an originator may keep inquiring about a specific borrower's information for multiple times;
2) an originator may inquire multiple borrowers within a single group to get more information of that group.
This means that the change of the loan information between a borrower and a lender affects not only the borrower's type counts but also other borrowers' type vectors.
Actually, if we assume that an originator would repeat the query on a borrower for at most $k$ times, then the worst case would be $\vec{v} = (\vec{v}^{k}_1, \vec{v}^{k}_2, \dots, \vec{v}^{k}_m)$. 
Futhermore, if there are at most $l$ borrowers whose type vectors may be affected by a specific borrower, then at most $lk$ queries of an originator would be affected by a specific borrower.
To still achieve $(\epsilon, \delta)$-differential privacy, we can then split the privacy budget $\epsilon$ and $\delta$ equally to the $lk$ queries.
However, such split scheme makes the average and standard variance of the amount of the noise that follows $\lceil \max(0, \lap{\mu, \lambda}) \rceil$ grow linearly with $lk$. We thus should reduce $l$ to improve the overall performance. 
Specifically, we have the following theorem (proof in Appendix~\ref{appendix:proof_dp_general}):

\begin{thm}
\label{thm:differential_privacy_general}
If an originator may inquire $m$ borrowers using Algorithm~\ref{algorithm:pir_sparse_recursive} with $d$-dimensional PIR queries and would inquire each borrower for at most $k$ times, then we have: \\
a) $l \le m^{(d-1)/d}$; \\
    b) if we replace all the empty items with $E^{d-s}(0_{s})$'s after the $s$-th iteration ($1 \le s \le d$), then $l \le m^{(s-1)/d}$ and $\hat{d} = s$.
\end{thm}

We can see from b) in Theorem~\ref{thm:differential_privacy_general} that the smaller $s$ is, the more computation the lenders take, but the fewer noise responses are needs as the privacy budgets $\epsilon$ and $\delta$ become larger and $\hat{d}$ becomes smaller. 
This is because the empty items are replaced with $E^{d-s}(0_{s})$'s at the $s$-th iteration and thus cannot be skipped, meanwhile the number of noise types the exchanger generates grows with $s$. 
In this paper we call $s$ the \emph{replace iteration}.
One notable point is that if we set $s = 1$, then $l \le 1$, which means that we only need to take the inquired borrower himself into account, which reduces to Theorem~\ref{thm:differential_privacy}.


\para{PRF for generating randomness.}
We use a PRF to generate randomness for the commitments. 
As mentioned in Section~\ref{section:architecture}, as long as $u$ has borrowed money from $S_i$, they share a secret string $\tau_{iu}$ and a number $x_{iu}$. 
The PRF uses $\tau_{iu}$ as seed and takes a string containing $x_{iu}$ as input, and outputs the randomness $r_{iu}$. Thus $u$ and $S_i$ can privately share the randomness without communication. 
We summarize our secure aggregation protocol (denoted by $\Pi_{agg}$) in Appendix~\ref{appendix:pir_sparse_recursive}.

\section{Process 2: Anonymous Authorization}
\label{section:anonymous_authorization}
The above protocol is enough for a semi-honest originator to aggregate the commitments if the originator follows the protocol. 
However, the originator may deviate from the protocol by faking queries that violate \emph{Privacy-L} and \emph{Authorization}.
There are two ways that the originator can cheat the exchanger and the lenders: invalid queries and unauthorized queries.

\para{Non-interactive ZKP for query validity.}
We define a recursive PIR query as valid iff the subquery of each dimension $i$, i.e. $\{q_{i1}, q_{i2}, \dots, q_{im_i}\}$, contains exactly one $1$ and $m_i-1$ $0$'s. 
As the query is encrypted, a malicious originator can send a query with more $1$'s  to retrieve information on other borrowers.
To avoid such an attack, we use ZKP to prove two constraints for each dimension $i$ of the query:
a) each $q_{ij} (j = 1, 2, \dots, m_i)$ encrypts either $0$ or $1$; 
and b) $\prod_j^{m_i} q_{ij}$ encrypts 1. 
Appendix~\ref{appendix:valid_query} provides the formal description and proof.
It is easy to verify that these two constraints indicate that the subquery of dimension $i$ is valid, thus the entire query.
We also use the \emph{Fiat-Shamir heuristic}~\cite{FS_HEURISTIC} 
to make the above proof non-interactive.
In \sysname, the exchanger verifies the proofs and rejects the query if the verification fails.

\para{Authorization the query anonymously.  }
Another subsistent issue is unauthorized queries: a malicious originator can inquire loan information of a borrower $b$ without her permission. 
For example, the originator can first find a pretender $b'$ to perform as $b$. 
Then the originator initiates a query about $b$. 
As $b'$ communicates with the exchanger anonymously, no one except the originator knows the identity of $b'$.
The originator, though not able to open the commitments from the lenders, can deduce how many lenders $b$ has borrowed money from as follows: 
As steps $e$ and $f$ in Protocol~\ref{protocol:pir_sparse_recursive} show, the borrower sends $\Delta r_b$ to the exchanger and the exchanger sends $\Delta r = \Delta r_b - r_z$ to the originator, thus if the pretender $b'$ colludes with the originator, the originator actually knows the value of $\Delta r_b$ and can recover $r_z$ as $r_z = \Delta r_b - \Delta r$. 
Also, as the noise commitments all commit to $0$, with the set $C$ got in step $g$, the originator can find a subset $C_{sub} \subset C$ such that $r_z$ opens $\prod_{c_i \in C_{sub} c_i}$ to $0$. Finally, the originator gets to know that there are $\big| C - C_{sub} \big|$ lenders who have lent money to $b$. 

Thus, our goal is to make sure that the exchanger knows that the ``borrower'' communicating with it is actually the borrower that the originator is inquiring about. 
Existing proving schemes for identity, such as \cite{ZKP_IDENTITY, ZKP_IDENTITY1, ZKP_IDENTITY2}, require the prover to reveal her identity to the verifier. 
Also, membership ZKP schemes, such as \cite{RING_SIGNATURE, RING_SIGNATURE1, MEMBERSHIP_ZKP}, only prove that a player is one of the members in a specific group, but cannot distinguish $b$ and $b'$ when $b'$ is also a legit registered user in the same group. 
Our solution is intuitive: 
the exchanger first uses the PIR query on the dataset $T_E = \{\tau_{eu}\}$ where $\tau_{eu}$ is a secret number shared between the exchanger and a user $u$ in that group, to retrieve the current borrower's secret, and then asks the borrower and the originator to collaboratively prove that the anonymous borrower knows the retrieved secret. As the secret retrieved by the exchanger is encrypted, the borrower's identity is still kept secret from the exchanger.

Specifically, given the borrower's identity $b$, if the exchanger directly performs PIR using a $d$-dimensional query, it gets $\text{E}^d(\tau_{eb})$, and the borrower should prove that she knows the plaintext of $\text{E}^d(\tau_{eb})$ without revealing her identity. 
When $d = 1$, in the Paillier encryption scheme, the proving process can be accomplished using the approach in \cite{PAILLIER_ZKP}. 
When $d > 1$, however, it is non-trivial to construct an efficient proof for the statement. 
We thus propose an efficient new protocol for proving the knowledge of the plaintext of a ciphertext generated by a recursive PIR query. 
While the exchanger issuing PIR queries along each dimension, 
we require the borrower and the originator to convince the exchanger that for dimension $i (1 \le i \le d)$, the secret the borrower knows, $\tau_{eb}$,  is in the items retrieved by the subquery of dimension $i$.
In this situation, the ciphertexts produced by each subquery have only one layer of encryption, thus we can use existing cryptographic tools to prove the above claim.
Here ``performing PIR along each dimension respectively'' means that we treat an array of $m$ items as a $d$-dimensional array with shape $m_1 \times m_2 \times \dots \times m_d$, and for each dimension $i$, we use the subquery of that dimension to retrieve $m / m_i$ items.
If the borrower really knows a secret number that can be retrieved by every subquery, then we can conclude that the secret she knows is exactly the item the recursive PIR query retrieves, as no other items are in the intersection of all the subqueries.
Appendix~\ref{appendix:correspondence} provides the formal description and analysis of the protocol. 

\begin{figure}[tb]
    \centering
    \includegraphics[width=0.30\textwidth]{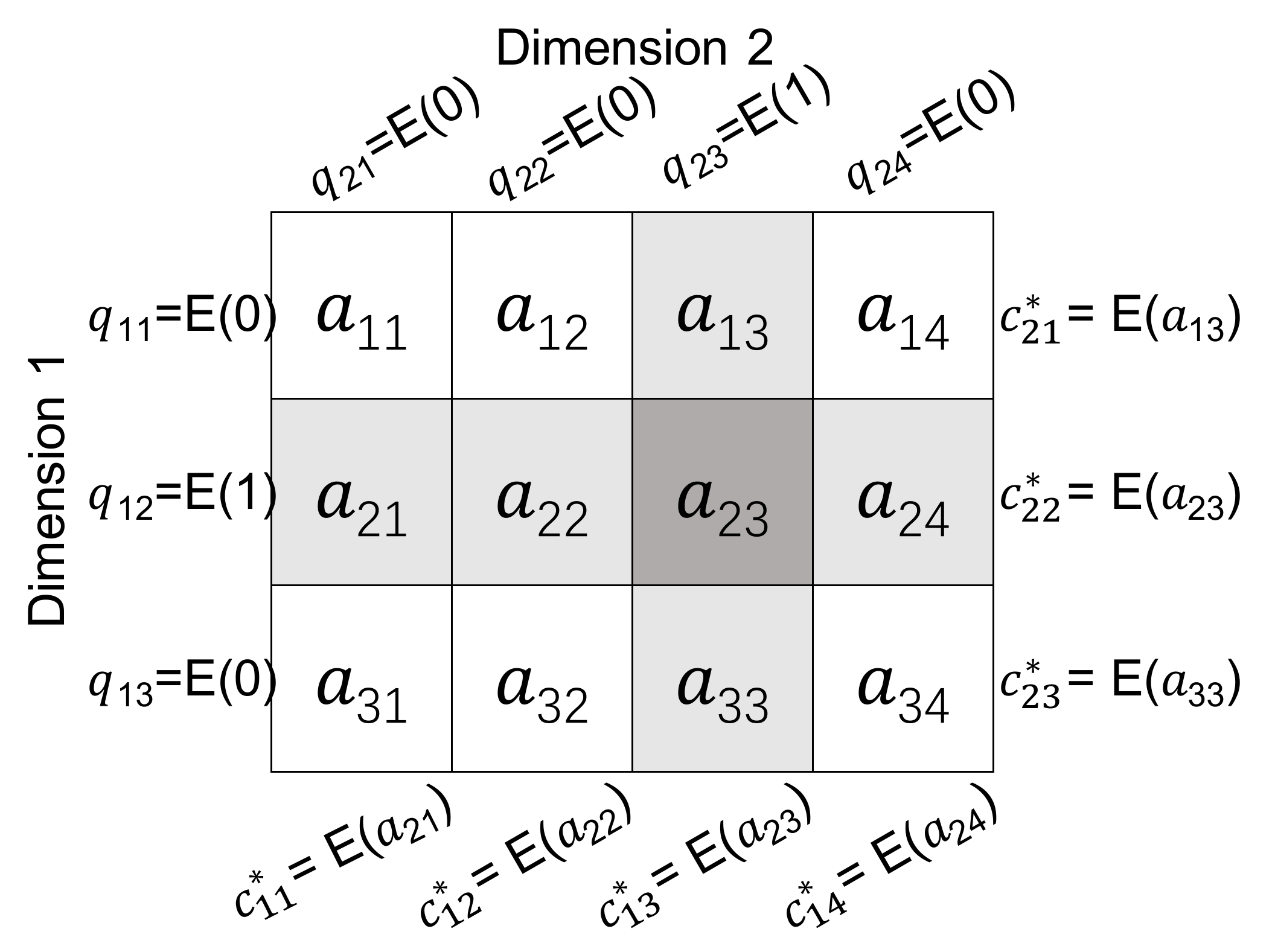}
	\vspace{-0.2in}
    \caption{An example of anonymous authorization.}
	\vspace{-0.1in}
    \label{figure:identity}
\end{figure}

Fig.~\ref{figure:identity} uses a $2$-D query on a $3 \times 4$ array $A$ to illustrate the process.
The borrower, who holds $a_{23}$, first sends a ciphertext $c = \enc{a_{23}}$ to the exchanger along with the corresponding proof that she knows the plaintext of $c$. Then for dimension $1$, the exchanger uses the subquery $\{q_{11}, q_{12}, q_{13}\}$ and gets $(c^*_{11}, c^*_{12}, c^*_{13}, c^*_{14})$, which encrypts the 2nd row. And as $a_{23}$ is in the 2nd row, the borrower can prove to the exchanger that the secret encrypted in $c$ is also encrypted in one of $c^*_{11}, c^*_{12}, c^*_{13}, c^*_{14}$.
Similarly, for dimension $2$, the exchanger retrieves the 3rd column and the borrower proves that one of $c^*_{21}, c^*_{22}, c^*_{23}$ encrypts the same number as $c$.
Thus, the secret encrypted in $c$ is in both the 2nd row and the 3rd column, which indicates that the borrower knows $a_{23}$.


\para{Newly generated random numbers. } 
Instead of using $\tau_{eu}$'s directly for PIR, we use $\tau_{eu}$'s as seeds to generate new pseudo-random numbers for each query to prevent the leakage of $\tau_{eu}$'s.
We formalize the anonymous authorization protocol (denoted by $\Pi_{auth}$) in Appendix~\ref{appendix:anonymous_authorization}.


\section{Process 3: Secure Evaluation}
\label{section:secure_evaluation}

The third process is to enable the borrower and the originator to securely evaluate function $f$ in Eq.~\ref{equation:query}.
The output of $f$ is based on the aggregation of $x_{ib}$, but unlike some existing work such as \cite{PRIO, PEM} which reveal the aggregation results (i.e. $\sum_i x_{ib}$) directly, \sysname also supports further operations on $\sum_i x_{ib}$ and the originator's private data $t$ without revealing $\sum_i x_{ib}$ and $t$.
We use ZKP and \emph{non-interactive actively secure computation (NISC)}~\cite{NISC} to achieve this goal. 
Both ZKP and NISC is based on the commitment that the borrower has sent to the originator for consistency check at the beginning of secure aggregation,
Note that, although all the communication between the borrower and the originator is through the exchanger, we omit the exchanger in this section for simplicity, as the exchanger only relays messages.

\sysname supports several kinds of queries, including 1) sum-based queries such as \code{sum} and \code{count}; 2) queries containing multiplication gates, such as \code{variance}; and 3) non-linear functions such as \code{comparison}.  We briefly introduce how we implement these queries.

\para{Sum. }
The implementation for the sum query is direct: the borrower only needs to open the aggregated commitment in Process 1 and reveals $\sum_i x_{ib}$ to the originator.

\para{Count. }
The count query tells the originator how many lenders have lent money to the borrower.  It can be computed by replacing the loan amount with 1 if $x_{ib} > 0$, or $0$ otherwise.

\para{Variance. }
The variance of $x_{ib}$'s can be calculated as $(\frac{1}{n}\sum_i x_{ib}^2) - (\frac{1}{n}\sum_i x_{ib})^2$.
Intuitively, the originator can first get the commitment to $\sum_i x_{ib}^2$ (denoted by $F_1$) and the commitment to $(\sum_i x_{ib})^2$ (denoted by $F_2$), then the originator calculates $F_3 = F_1^n F_2^{-1}$, which is the commitment to $(n\sum_i x_{ib}^2) - (\sum_i x_{ib})^2$, and opens $F_3$ with the help of the borrower.
The originator can obtain $F_1$ by collecting the commitments to $x_{ib}^2$'s. However, it is tricky to get $F_2$ using the commitments from the lenders, as the commitment scheme is not multiplicatively homomorphic. 
\cite{PRIO} calculates the variance by revealing $(\frac{1}{n}\sum_i x_{ib}^2)$ and $(\frac{1}{n}\sum_i x_{ib})^2$ directly, but it leaks extra information beyond the variance.
\sysname uses the \emph{multiplication ZKP}~\cite{ZKP_HOMO} that proves that a committed number is the product of the two numbers in another two commitments, and thus only reveals the final variance.
Specifically, to get the commitment to $(\sum_i x_{ib})^2$, the originator first uses secure aggregation to retrieve the commitment to $\sum_i x_{ib}$ (denoted as $F_4$), while the borrower sends $F_2$ (i.e. the commitment to $(\sum_i x_{ib})^2$) along with the proof which proves to the originator that the number committed in $F_2$ is the square of the number committed in $F_4$. Once the originator verifies that $F_2$ is the commitment to $(\sum_i x_{ib})^2$, it can obtain the variance using $F_1$ and $F_2$.

\para{Comparison to a public number. }
Consider a case where the originator is willing to make the credit limit $t$ public. Then $f$ is the comparison function which evaluates whether $\sum_{i=1}^{n} x_{ib} > t$.
We use the \emph{interval ZKP}~\cite{INTERVAL_ZKP}, which proves that a committed number lies in a public interval.
The interval ZKP is efficient and non-interactive. In \sysname, the borrower sends the proof to the originator and then the originator verifies the proof to get the comparison result.

\para{Comparison to a private number. }
In a more realistic setting, the originator also wants to hide $t$.  
If we assume semi-honest borrower and originator, we can directly apply \emph{garbled circuit (GC)}~\cite{GC} to perform the comparison. However, both the borrower and the originator have the incentive to deviate from the protocol and to lie. We thus use the \emph{non-interactive actively secure computation (NISC)} scheme in \cite{NISC} to support such queries.
The NISC scheme provides active security for GC using \emph{cut-and-choose}~\cite{CUT_AND_CHOOSE}. Also, we employ common optimizations for GC, such as \emph{free-xor}~\cite{FREE_XOR} and \emph{half-and}~\cite{TWO_HALVES}.
We make a small modification: although the protocol in \cite{NISC} generates input commitments to prove the consistency of inputs of different circuits, in our scenario, however, the originator needs to make sure that the input commitments for secure computation are also consistent with the commitment in the consistency check process, otherwise the borrower is still able to use a fake input for secure comparison to cheat the originator. 
Thus, in \sysname, the borrower also needs to prove consistency. 
Specifically, in \cite{NISC}, the circuit generator (namely the borrower in this paper) generates an input commitment $c_j$ for the $j$-th bit of her input.
Then the borrower in \sysname generate proofs for two constraints:
a) each $c_j$ commits either $0$ or $1$;
b) $c = \prod c_j^{2^j}$ commits the same number as the commitment in the aggregation process. 
The borrower sends the proofs to the originator through the exchanger, and the originator verifies them to get the comparison result.

\section{Security Analysis}
\label{sec:security_analysis}

For the formal security analysis, we construct simulators for the participants and prove the indistinguishability between the real view and the simulated view in each single subprocol, then analyze the security of the composition of the subprotocols. 
Briefly speaking, we first see that in $\Pi_{auth}$, the participants only receives zero-knowledge proofs, commitments, ciphertexts or random numbers, which reveal no information about the others' original inputs and can be simulated by the simulators, thus achieving the indistinguishability. And the zero-knowledge proofs help to detect malicious originators and borrowers.
We then prove that with the functionality of $\Pi_{auth}$, $\Pi_{agg}$ also provides security and uses the commitment scheme to detect malicious borrowers.
Finally, with the composition of $\Pi_{auth}$ and $\Pi_{agg}$, we show the security of the whole protocol.
We further argue that we achieve the security goals in Section~\ref{section:security_goal}.
Please see Appendix~\ref{appendix:formal_proof} for details.

In addition, we consider an adversary who controls all the connections and monitors all the traffic in \sysname.
As we assume that all the communications use secure channels (e.g. SSL) and all transferred data is in encrypted or committed version, the adversary cannot infer any information by observing the messages or the sizes of messages in the channels. 
Another possible attack for the adversary is to block connections and observe the result. However, blocking the borrower or the originator makes no sense, as the protocol aborts if either of them is blocked and there would be no observable plain variables for the adversary. 
Blocking one or more lenders does not hurt the security of \sysname either, as the exchanger perturbs the responses from the lenders using noise before the responses are decrypted.
So we can conclude that \sysname is also secure against such adversaries.

A notable thing is that, the computation overhead of each lender is proportional to the number of its registered borrowers. One may argue that such design is vulnerable to timing attack~\cite{TIMING, TIMING_MIX}. But we do not think this as a problem in \sysname, due to the following reasons:
1) The lenders can wait for a random period of time before sending data out to prevent the adversaries from capturing the relation between time and count;
2) Different lenders use different infrastructure to perform the computation, thus longer time does not necessarily mean more non-empty items.
3) The rough number of users of a lender sometimes is not privacy, as public materials such as financial reports may also reveal such information.

\section{Evaluation}
\label{section:evaluation}

\subsection{Implementation and Testbed Setup}
\label{section:setup}

We implement \sysname prototype with about $4,200$ lines of C++ code. 
We use \emph{OpenMP}~\cite{OPENMP} to parallelize the computation. 
To support big integers, we use the \emph{GMP} library~\cite{GMP}. We use the \emph{Crypto++} library~\cite{CRYPTOPP} for common cryptographic tools, such as \code{SHA256} and \code{AES}.

We use the method in \cite{APPLIED_CRYPTOGRAPHY} to generate the parameters $g$ and $h$ for the commitment function, 
while using the method in \cite{RANDOMNESS} to implement PRFs.
Also, we use $1024$-bit Paillier for PIR.
To reduce the time for online computation, the exchanger generates noise responses offline for different originators.
We use the secure socket layer (SSL) to add security and use Tor~\cite{TOR} to add anonymity to the communication channels between the borrower and exchanger.

Finally, we run our experiments on Amazon EC2 virtual machines. All nodes are of type \code{c5.2xlarge} with 8 Intel Xeon Platinum 8000-series CPU cores and 16 GB RAM. Each node runs 64-bit Ubuntu 16.04 with 4.4.0 kernel.
Each role of \sysname runs in a separate EC2 node in our experiments. To evaluate the scalability of \sysname, we scale the number of lenders up to 800, which means 803 virtual machines in total (800 lenders + 1 borrower + 1 originator + 1 exchanger).

\subsection{Configuration parameters.  }
There are many configuration parameters to set in \sysname's processes 1-3.  We first summarize them here and we make comprehensive evaluations on their effects in this section.

\para{PIR parameters.  }
Recall that we perform recursive PIR on a group of $N_g$ members.  
We set $N_g = 10,000$ in this paper, so that the lenders can only distinguish the current borrower from other registered users with probability $1/N_g = 10^{-4}$, which is good enough in most cases. 
The other essential PIR parameter is the dimension $d$ of a recursive PIR query. 
Specifically, we can treat a dataset of $N_g$ items as a $m_1 \times m_2 \times \dots \times m_d$ array and perform PIR $d$ times recursively.  In the evaluation, we choose two settings of $d$: 
a) $d = 4$ and $N_g = 10 \times 10 \times 10 \times 10$; 
and b) $d = 2$ and $N_g = 100 \times 100$.
To be succinct, we denote the two settings as $Q_{10}^{4}$ and $Q_{100}^{2}$, respectively.  
$Q_{10}^{4}$ generates 40 ciphertexts for each query with four sub-queries, while $Q_{100}^{2}$ generates 200 ciphertexts for each query with two sub-queries. 
Taking both computation and communication cost into account, we will show in our experiments that it is a trade-off to choose a proper kind of recursion.
We set the \emph{replace iteration} $s$ to 1 by default (see Theorem~\ref{thm:differential_privacy_general}).
We evaluate its impact in Section~\ref{section:evaluation_parameters}.

\para{Sparsity of a lender's database.  }
A considerable factor is the sparsity of the lenders' databases. As mentioned in Section~\ref{section:dspir}, each lender only needs to store the information of its own borrowers and performs PIR on the sparse dataset, skipping empty items.
Therefore, the sparsity strongly affects the performance of Algorithm~\ref{algorithm:pir_sparse_recursive} running on each lender.
The default value in our evaluation is 0.1, a common estimate~\cite{P2P_PERFORMANCE}.

\para{Differential privacy parameters.  }
The privacy budget for noise generation is another considerable factor. The goal of noise generation in \sysname is to achieve $(\epsilon, \delta)$-differential privacy after $k$ rounds of a query. 
By default, we set $\epsilon = ln2 \approx 0.7$ and $\delta = 10^{-4}$ as recommended in \cite{VUVUZELA}.
We also set $k=5$, which we think sufficient and practical in most cases, as we present in Section~\ref{section:dspir}.

\para{Network configurations.  }
As the network is usually the bottleneck, we compare the following three network settings. 

\noindent [\emph{WAN}] To simulate the typical Internet settings, we limit the bandwidth for all nodes to 8 Mbps (i.e. 1 MB/sec) and the latency of each node to 50ms.

\noindent [\emph{LAN}] is the raw EC2 network setting. The bandwidth among nodes ranges from tens of Mbps to 10 Gbps. 
We consider this configuration because the lenders may be a tight coupling consortium (e.g. most banks in China) and have their servers in co-located data centers, which make them enjoy LAN speed.

\noindent [\emph{EXC}]  As it is likely that the bandwidth of the exchanger becomes the bottleneck, we provide a more realistic setup where exchanger has 800 Mbps bandwidth while other nodes' bandwidth stays 8 Mbps. The network latency stays 50ms.

\subsection{Overall Performance}
\label{section:overall_performance}

\begin{figure}[tb]
	\centering
	\includegraphics[width = 0.5\textwidth]{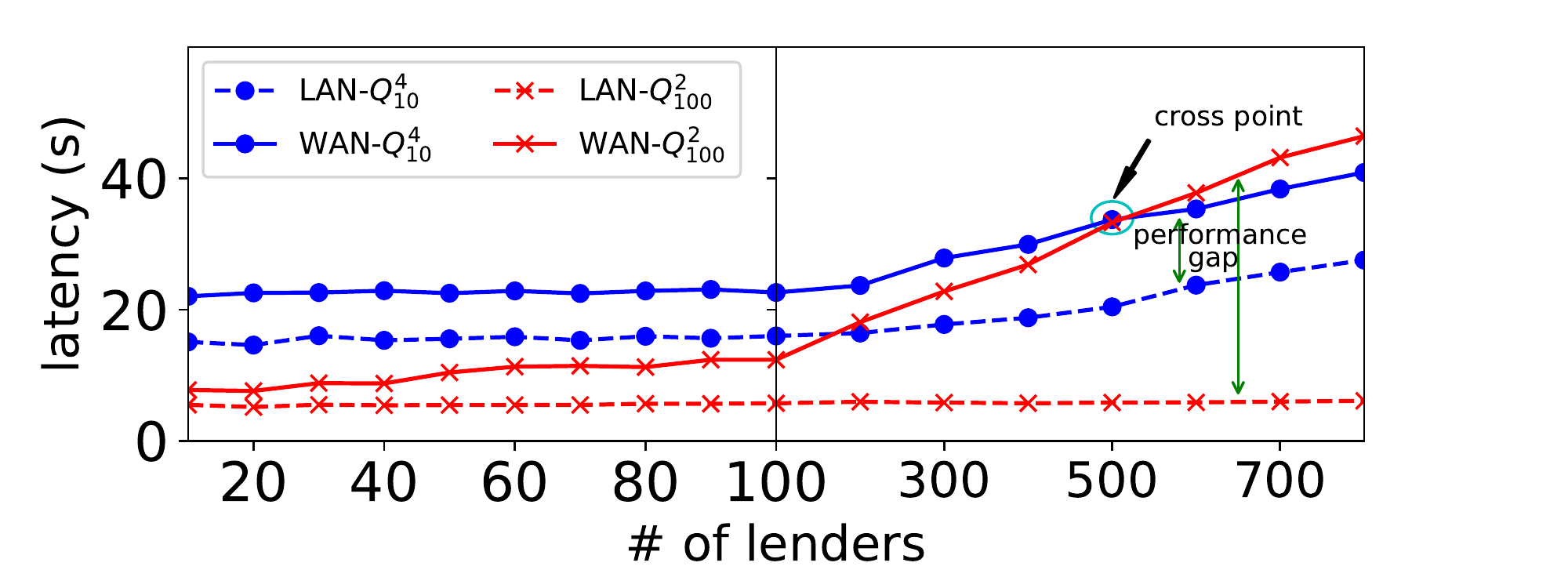}
	\vspace{-0.3in}
    \caption{Latency of \sysname in WAN and LAN.}
	\label{figure:latency_wan_1m}
\end{figure}

\begin{figure}[tb]
	\centering
	\includegraphics[width = 0.5\textwidth]{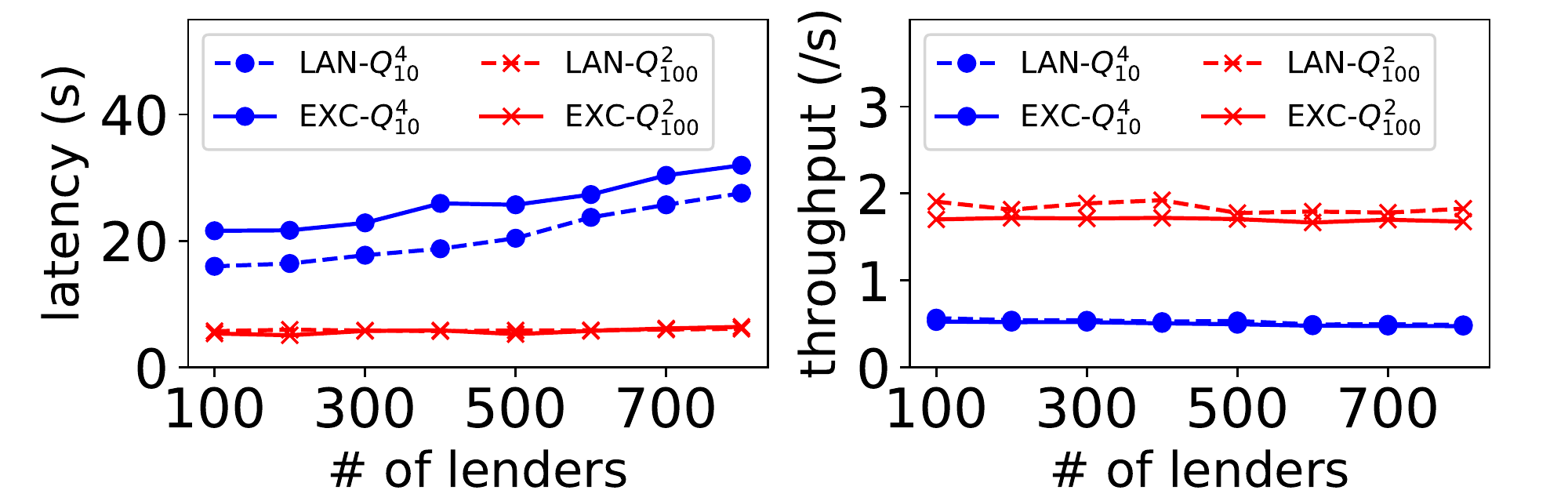}
	\vspace{-0.2in}
	\caption{Performance of \sysname in EXC and LAN.}
	\vspace{-0.1in}
	\label{figure:latency_throughput_wan_100m}
\end{figure}

We first present the overall performance of \sysname in the 800-node EC2 testbed with different network configurations.  We focus on introducing the performance in our default setting presented above under different numbers of lenders and network configurations, and we leave comparison to the other parameters in the following sections.  


\para{\emph{[WAN]} vs. \emph{[LAN]} performance.  }
We first evaluate the end-to-end latency of a query under both network settings. Fig.~\ref{figure:latency_wan_1m} shows the results, and we observe the following: 

\noindent 1) With a small number of lenders (i.e. $n < 100$), the whole process only takes several seconds (e.g. for $n=50$, WAN-$Q_{10}^{4}$:22.5s, WAN-$Q_{100}^{2}$:10.4s, WAN-$Q_{10}^{4}$:15.6s, WAN-$Q_{100}^{2}$:5.4s), showing that \sysname is practical even when the network resource is strictly limited.

\noindent 2) When the number of lenders $n$ is small, the overall performance is roughly independent of $n$. This is because the bottleneck is the computation in anonymous authorization. However, when $n$ gets large, the network becomes the bottleneck, and the total time increases linearly with $n$.

\noindent 3) The performance gap between LAN and WAN is large, meaning that the low bandwidth of 8 Mbps significantly limits the performance. 

The above evaluation also indicates that the size of a query, or communication cost, can be the bottleneck in WANs, especially with a large number of the lenders. 

\para{\emph{[EXC]} vs. \emph{[LAN]} performance.  } 
It is easy to see that the exchanger needs much larger bandwidth, and thus in the \emph{[EXC]} we increase the exchanger bandwidth to 800 Mbps.  Fig.~\ref{figure:latency_throughput_wan_100m} compares the performance between \emph{[LAN]} and \emph{[EXC]}.  
We can see that when the number of lenders increases, both the performance gap and the performance degradation become smaller, especially that the effect of the network nearly disapears for the $Q_{100}^{2}$ version.
Meanwhile, the effect of the network remains for the $Q_{10}^{4}$ version due to the bandwidth limits between the originator and the exchanger.

In addition to latency, we also plot the throughput in Fig.~\ref{figure:latency_throughput_wan_100m}.  We can see that \sysname can handle 2 $Q_{100}^{2}$ queries or 0.5 $Q_{10}^{4}$ queries per second. This result is practical enough for loan stacking detection. Actually, a query in existing production systems (without security) usually crosses multiple organizations (e.g. lenders), and the latency is determined by the slowest responding node. Also as the system should handle distribution issues such as connection fault tolerance and access control, it is normal to take several seconds and done in an asynchronous manner.

\subsection{Cost Breakdown}

\para{Protocol 1 and 2.  } Table~\ref{table:secure_aggregation_time} summarizes the time consumed in each step in secure aggregation and anonymous authorization, while Table~\ref{table:secure_aggregation_size} shows the size of data generated and transferred by each role. 
We use the initials as shorthand for a role, i.e., $o$, $b$, $e$ and $l$ stand for originator, borrower, exchanger, and lender, respectively.
From the results, we can see that, for both computation and communication, anonymous authorization is the main bottleneck. But the overall performance is practical, as the whole computation only takes several seconds.
We also compare the results of the two kinds of recursions. We can see that, in most steps, a $Q_{100}^{2}$ query outperforms a $Q_{10}^{4}$ query. However, in the aspect of generating a PIR query with proof, $Q_{10}^{4}$ outperforms $Q_{100}^{2}$ both in terms of computation time and data size, as expected. 

\begin{table}[tb]
	\footnotesize
	\renewcommand{\arraystretch}{1.3}
	\begin{center}
		\begin{tabular}{c l c c}
			\hline
			\multirow{2}{*}{role} & \multicolumn{1}{c}{\multirow{2}{*}{action}} & \multicolumn{2}{c}{time (s)} \\
			\cmidrule(lr){3-4}
			& & $Q_{10}^{4}$ & $Q_{100}^{2}$ \\
			\hline
			$o$ & \tabincell{l}{generate query and proof of right form} & 0.083 & 0.306 \\
			$e$ & \tabincell{l}{verify proof is well-formed } & 0.067 & 0.239 \\
            $l$ & \tabincell{l}{perform PIR using Alg.~\ref{algorithm:pir_sparse_recursive} (sparsity = $0.1$, $s = 1$)} & 1.80 & 0.62 \\ 
            $e$ & \tabincell{l}{generate noise (offline, $\epsilon = 0.7, \delta = 10^{-4}, k = 5$)} & & \\ 
			$o$ & \tabincell{l}{decrypt each response} & 0.041 & 0.006 \\
			\hline
			$e$ & \tabincell{l}{generate data for authorization} & 2.11 & 1.05 \\ 
			$o$ & \tabincell{l}{generate proof for the borrower's identity} & 4.45 & 1.20 \\ 
			$e$ & \tabincell{l}{verify proof for the borrower's identity} & 2.28 & 0.118 \\ 
			\Xhline{1.5\arrayrulewidth}
		\end{tabular}
		\caption{Computation time of each step in secure aggregation and anonymous authorization.}
		\label{table:secure_aggregation_time}
	\end{center}
\end{table}

\begin{table}[tb]
	\footnotesize
	\renewcommand{\arraystretch}{1.3}
	\begin{center}
		\begin{tabular}{c l c c}
			\hline
			\multirow{2}{*}{generator} & \multicolumn{1}{c}{\multirow{2}{*}{description}} & \multicolumn{2}{c}{size (KB)} \\
			\cmidrule(lr){3-4}
			& & $Q_{10}^{4}$ & $Q_{100}^{2}$ \\
			\hline
			$o$ & \tabincell{l}{the PIR query} & 10.4 & 51.5 \\
			$o$ & \tabincell{l}{the proof of the query} & 54.1 & 259 \\
			$l$ & \tabincell{l}{a PIR response} & 3.86 & 0.771 \\
            $e$ & \tabincell{l}{average noise responses (offline)} & 1191.5 & 250.6 \\ 
			\hline
			$e$ & \tabincell{l}{array for authorization} & 1370 & 1370 \\
			$o$ & \tabincell{l}{the proof of the borrower's identity} & 2061 & 104 \\
			\Xhline{1.5\arrayrulewidth}
		\end{tabular}
		\caption{Size of data generated in secure aggregation and anonymous authorization.}
		\label{table:secure_aggregation_size}
	\end{center}
\end{table}

\para{Protocol 3. }  
Table~\ref{table:secure_evaluation_time} shows the running time and the transferred data size of different kinds of queries in Process 3 (secure evaluation). 
\emph{sum} and \emph{count} come directly from the secure aggregation, thus we can get them for free. 
For the \emph{variance} and \emph{comparison with public} queries, as we employ non-interactive ZKP, the data sent by the originator is 0 KB. 
The more complex queries such as \emph{comparison with private} use NISC for secure evaluation, requiring the borrower to generate and send out multiple encrypted circuits.  However, it is still KBs in size, much smaller than the previous processes. 

\begin{table}[tb]
	\footnotesize
	\renewcommand{\arraystretch}{1.3}
	\begin{center}
		\begin{tabular}{c c c c}
            \hline
			role & variance & \tabincell{c}{cmp with public} & \tabincell{c}{cmp with private} \\
			\hline
			$b$ & 0.0018s, 0.75KB & 0.009s, 2.69KB & 0.143s, 1120KB \\
			$o$ & 0.002s, 0KB & 0.011s, 0KB & 0.258s, 18.4KB \\
			\Xhline{1.5\arrayrulewidth}
		\end{tabular}
		\caption{Time and data size in secure evaluation.}
		\label{table:secure_evaluation_time}
	\end{center}
\end{table}

\subsection{Parameters for Process 1 and 2}
\label{section:evaluation_parameters}
Process 1 and 2 (\emph{secure aggregation} and \emph{anonymous authorization}) involve most computation time, and thus we would like to further evaluate factors that affect the performance.  
To better illustrate the effects of the parameters, we extend the notation $Q_{n_i}^d$ to $Q_{n_i}^{d,s}$, where $s$ is the replace iteration.

\begin{figure}[tb]
    \centering
    \includegraphics[width = 0.33\textwidth]{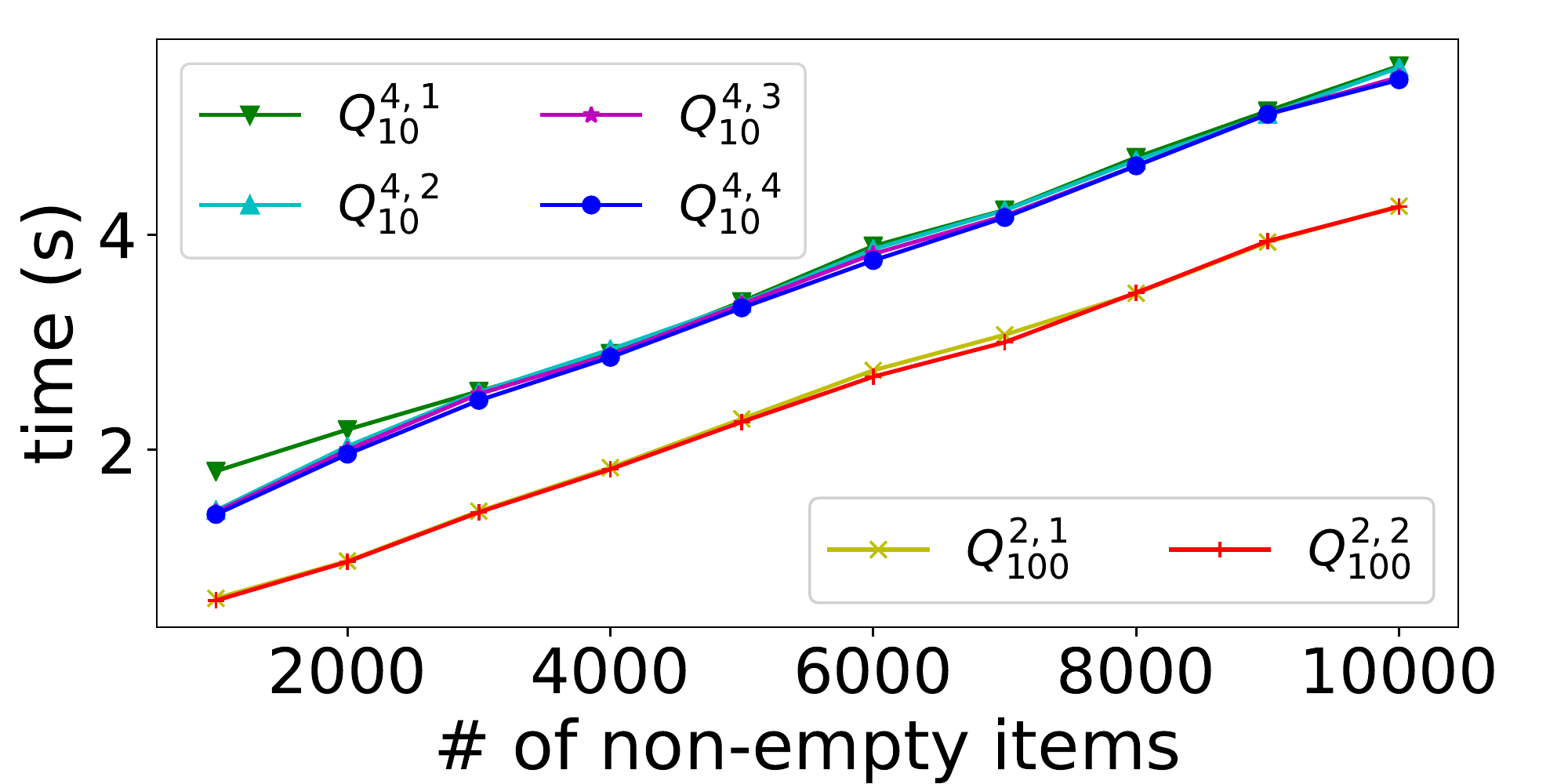}
    \vspace{-0.15in}
    \caption{The performance of Algorithm~\ref{algorithm:pir_sparse_recursive}.}
    \vspace{-0.1in}
    \label{figure:time_pir}
\end{figure}

\para{Sparsity. }
Fig.~\ref{figure:time_pir} shows the relation between the sparsity and efficiency. 
The $Q_{100}^{2,s}$ queries perform better than the $Q_{10}^{4,s}$ ones. The reason is that fewer dimensions mean fewer recursions, leading to less computation cost.
Another notable thing is that the effect of the replace iteration $s$ is small, especially for small sparsities. This is because the probability of a column being totally empty is small and there are not many $0$'s to replace, and thus the computation cost is low. 

\begin{figure}
    \centering
    \includegraphics[width = 0.45\textwidth]{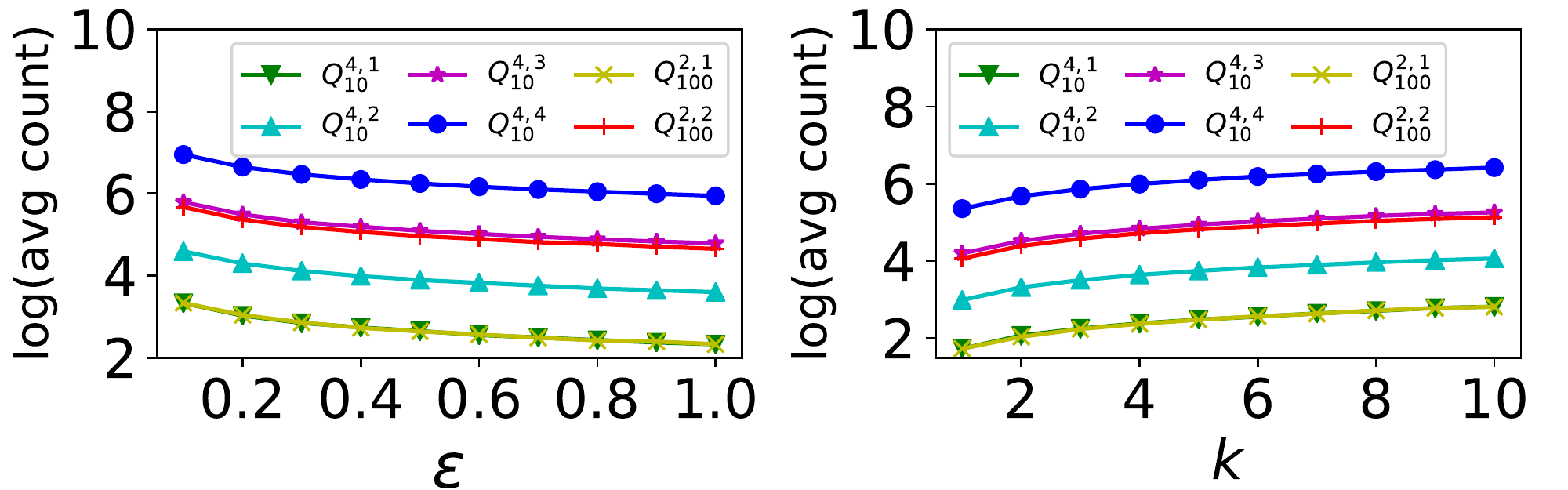}
    \vspace{-0.15in}
    \caption{The average number of noise responses.}
    \label{figure:noise_count}
\end{figure}

\para{Differential privacy parameters .  }
Differential privacy does come with a cost, and its parameters affect the noise size. 
We evaluate the average number of generated noise responses for different values of $\epsilon$ and $k$.
As Fig.~\ref{figure:noise_count} shows, a smaller $\epsilon$ provides stronger privacy, but leads to more noise.
We also vary the value of $k$ and perform the same evaluation.
Similarly, larger $k$ allows more chances of inquiring about a specific borrower, but requires more noise to prevent accumulated privacy leak.
From the figure, we can see that the time for generating many noise responses is non-trivial. 
Luckily, the noise is independent of the queries and the commitments, so we can pre-generate these noise responses offline, and thus still keep the online part fast. 

\para{Replace iteration $s$. }
To choose a proper $s$, we first see from Fig.~\ref{figure:noise_count} that if we set the replace iteration $s \ge 2$, the number of noise responses grows rapidly and would bring much larger ($20\times \sim 200\times$) communication cost than the case where $s = 1$. 
On the other hand, Fig.~\ref{figure:time_pir} shows that the replace iteration $s$ does not affect the sparse PIR time much. Thus we set $s=1$ in our evaluation.  

\para{Recursive PIR parameter. }
Both secure aggregation and anonymous authorization involve PIR queries.
Generally speaking, the $Q_{100}^{2}$ version outperforms the $Q_{10}^{4}$ version both in terms of computation and communication, except for the query size. Our evaluation also shows that the choice of the recursive dimension is a trade-off: more recursions means more computation and communication between the exchanger and the originator, while fewer recursions mean more communication between the exchanger and the lenders.

\section{Discussion}

\para{More on recursive PIR parameters.  }
The main computation cost comes from generating PIR responses and anonymous authorization. 
A higher-dimensional recursive PIR query means more full-database scans, and our evaluation also shows that $Q_{100}^{2}$ performs better than $Q_{10}^{4}$ in most settings.
However, fewer recursions does not always mean better performance. 
Let us consider a non-recursive query, i.e. $Q_{10000}^1$. It can be seen that such a PIR query using the Paillier cryptosystem with a $1024$-bit key is about $2.56$MB. And the proof size would be about $12.8$MB. Thus, the originator should send more than $15$MB to the exchanger, meaning that the latency of a query would be more than $15$s on a 8Mbps network. On the other hand, as the exchanger sends the query to all the lenders, the throughput would be less than $1 / 2.56 \approx 0.4$. Both the latency and throughput are worse than those of $Q_{100}^{2}$.

\para{FHE-based PIR. }
Systems like XPIR~\cite{XPIR} and SealPIR~\cite{SEALPIR} employ somewhat fully homomorphic encryption (FHE).  The advantage of FHE is much lower computation cost, due to avoiding modular exponentiation of large numbers. Unfortunately, the size of a ciphertext in this scheme is much larger than a ciphertext using Paillier, and thus we need to look for a trade-off. 
The size of a $d$-dimensional PIR query is at least $dN_g^{1/d}l$, where $l$ is the size of a ciphertext. 
And the size of the proof for the validation of a query is about $5 \times$ larger, and thus the originator should send at least $6dN_g^{1/d}l$.
On the other hand, the expansion factor of XPIR is $5$, which means that the size of each PIR response is about $5^d l$.
With $n$ lenders, there are at least $(6dN_g^{1/d} + n5^d)l$ bytes of  data transferred between the originator and the exchanger.
For $N_g = 10000$ and $n = 100$, we have $(6dN_g^{1/d} + n5^d)l \ge 3700l$.
With the default parameters in~\cite{XPIR}, the size of a ciphertext $l$, is about $64$KB. Therefore, there is at least $3700c = 236.8$MB data to be transferred between the originator and the exchanger, not practical in the wide-area network with many participants.  
SealPIR, though avoiding large queries, still suffering from large responses. For $d=2$ recommended in \cite{SEALPIR}, the size of a query is $64$KB, while the size of a response is $256$KB. 
Thus there are at least $6 \times 0.064 + n\times 0.256$ = $25.984$MB bytes of data to be transferred between the originator and exchanger.
Thus, we can see that FHE-based PIRs are not practical enough in our setting.

\section{Conclusion and Future Work}
\label{section:conclusion_future}
We propose \sysname, the first practical distributed system for privacy-preserving \emph{loan stacking detection}.
The process of \sysname includes three processes which can run in parallel: secure aggregation, anonymous authorization, and secure evaluation.
For secure aggregation, we propose a method to retrieve the commitments with differential privacy, and use zero-knowledge proofs to authorize the anonymous borrower's identity.
For secure evaluation, we support multiple kinds of efficient queries, including linear and non-linear ones. 
We then evaluate \sysname and show the trade-offs for selecting parameters. 
The evaluation demonstrates that \sysname can handle queries within a second in a real-world setting, and is practical for privacy-preserving credit evaluation.

As future work, we will add more features to \sysname. 
For example, we can use \emph{ring signature}~\cite{RING_SIGNATURE} to enable the originator to anonymously send an authorized query. 
Also, we can borrow the idea of \cite{ZEROCASH} to enable anonymous payment for each query.
Moreover, we can use distributed synchronization systems like \cite{ZOOKEEPER} for maintaining global configurations and status.
Last but not least, we will add fault tolerance mechanisms to make \sysname more robust.
We believe that with these improvements, our design would be more deployable in real-world scenarios.

\newpage

\bibliographystyle{plain}
\bibliography{reference.bib}

\newpage

\appendix

\section{Notations}
\label{appendix:notation}
Table~\ref{table:notation} summarizes the notations used in this paper. Readers can refer to this table for convenience.
\begin{table}
\center
	\begin{tabular}{|c|l|}
        \hline
        $n$ & \tabincell{l}{the number of lenders} \\
        \hline
        $S_i$ & \tabincell{l}{the $i$-th lender where $0 \le i < n$} \\
		\hline
		$x$ & \tabincell{l}{the total amount of money the current borrower \\ has borrowed} \\
		\hline
		$x_i$ & \tabincell{l}{the amount of money the current borrower has \\ borrowed from $S_i$} \\
		\hline
        $x_{iu}$ & \tabincell{l}{the amount of money the borrower with identity \\ $u$ has borrowed from $S_i$} \\
		\hline
		$\tau_{iu}$ & \tabincell{l}{the secret string of the borrower with identity $u$ \\ shared with $S_i$} \\
		\hline
		$\tau_{eu}$ & \tabincell{l}{the secret string of the borrower with identity $u$ \\ shared with the exchanger} \\
		\hline
		$d$ & \tabincell{l}{the dimension of a recursive PIR query} \\
		\hline
        $\hat{d}$ & \tabincell{l}{the number of noise types of $0$ string ciphertexts} \\
		\hline
        $s$ & \tabincell{l}{the replace iteration, after which \\ we replace all empty items with $E(0)$'s} \\
		\hline
		\tabincell{l}{$pk_o$ \\ $sk_o$} & \tabincell{l}{the public-private key pair of the originator} \\
		\hline
        \tabincell{l}{$\enc{\cdot}$} & \tabincell{l}{the public-key encryption scheme} \\
		\hline
        \tabincell{l}{$\text{E}^d(\cdot)$} & \tabincell{l}{applying the same public-key encryption scheme \\  for $d$ times} \\
		\hline
        \tabincell{l}{$0_l$} & \tabincell{l}{a 0 string which has the same length with $\text{E}^l(0)$} \\
		\hline
        \tabincell{l}{$Q_m^d$} & \tabincell{l}{a $d$-dimensional query for $m^d$ items} \\
		\hline
        \tabincell{l}{$Q_m^{d,s}$} & \tabincell{l}{a $d$-dimensional query for $m^d$ items \\ with replace iteration $s$} \\
		\hline
		\tabincell{l}{$p, q$ \\ $g, h$} & \tabincell{l}{the parameters of the commitment function} \\
		\hline
		$F(x, r)$ & \tabincell{l}{the commitment to $x$ with randomness $r$} \\
		\hline
		$N$ & \tabincell{l}{the number of registered users on the exchanger} \\
		\hline
		$N_g$ & \tabincell{l}{the number of registered users in each group} \\
		\hline
		$gid_u$ & \tabincell{l}{the group id of the user $u$} \\
		\hline
		$pid_u$ & \tabincell{l}{the position in the group of the user $u$} \\
		\hline
        $\epsilon, \delta$ & \tabincell{l}{the parameters of differential privacy} \\
		\hline
        $\mu, \lambda$ & \tabincell{l}{the parameters of Laplace distribution} \\
		\hline
	\end{tabular}
	\caption{The notations in this paper.}
	\label{table:notation}
\end{table}

\section{Proof of Theorem~\ref{thm:differential_privacy}}
\label{appendix:proof_dp}

Given the non-negative integer vector $\vec{v} = (n_0, n_1, \dots, n_d)$ where $n_i$ is the number of responses of $type \ i$, assume that a change of a lender's database results in a new vector $\vec{v}' = (n_0', n_1', \dots, n_d')$.
As each lender only sends one response to the exchanger, any change to a lender's database would only change the type of one response.
i.e., $\exists i,j \in \{0, 1, \dots, d\}, i \ne j$, $s.t.$
\begin{equation*}
    \begin{cases}
        n_i - n_i' = 1 \\
        n_j - n_j' = -1 \\
        \forall k \ne i,j \  n_k = n_k' \\
    \end{cases}
\end{equation*}

Now we fix the above $i$ and $j$, and consider a non-negative integer vector set $N'' = \{(n_0'', n_1'', \dots, n_d'') \ \big| \ n_i'' > \max(n_i, n_i'), n_j'' > \max(n_j, n_j')\}$. And we denote $\complement{N''}$ as the complementary set of $N''$.

As the sensitivity of $\vec{v}$ is $2$, i.e., the $L_1$-norm $\| \vec{v} - \vec{v}' \|_1 \le 2$, adding a noise vector $\widetilde{\vec{v}} = (\widetilde{n_0}, \widetilde{n_1}, \dots, \widetilde{n_d})$ achieves $\epsilon$-differential privacy, where $\widetilde{n_i} \sim \lap{\mu, \lambda}$ and $\epsilon = \frac{2}{\lambda}$,
i.e., for any set $S \subset N''$, we have $\pr{\vec{v} + \widetilde{\vec{v}} \in S} \le e^\epsilon \cdot \pr{\vec{v}' + \widetilde{\vec{v}} \in S}$.

Meanwhile, we calculate the probability that the perturbed vector is not in $N''$ as follows:
\begin{align*}
    & \ \ \ \ \ \pr{\vec{v} + \widetilde{\vec{v}} \not\in N''} \\
    &= \pr{n_i + \widetilde{n_i} \le \max(n_i, n_i') \text{ or } n_j + \widetilde{n_j} \le \max(n_j, n_j')} \\
    &= 1 - \\
    & \ \ \ (1 - \pr{n_i + \widetilde{n_i} \le \max(n_i, n_i')})(1 - \pr{n_j + \widetilde{n_j} \le \max(n_j, n_j')}) \\
    &\le 1 - (1 - \pr{n_i + \widetilde{n_i} \le n_i + 1})(1 - \pr{n_j + \widetilde{n_j} \le n_j + 1}) \\
    &= 1 - (1 - \pr{\widetilde{n_i} \le 1})(1 - \pr{\widetilde{n_j} \le 1}) \\
    &= 1 - (1 - \pr{\lap{\mu, \lambda} \le 1})(1 - \pr{\lap{\mu, \lambda} \le 1}) \\
    &= 1 - (1 - \frac{1}{2}e^\frac{1 - \mu}{\lambda})^2 \\
    &= e^{\frac{1-\mu}{\lambda}}(1- \frac{1}{4}e^{\frac{1-\mu}{\lambda}}) = \delta \\
\end{align*}

Therefore, for any non-negative integer vector set $S$, we have
\begin{align*}
    & \ \ \ \ \ \pr{\vec{v} + \widetilde{\vec{v}} \in S} \\
    &= \pr{\vec{v} + \widetilde{\vec{v}} \in S \cap N''} + \pr{\vec{v} + \widetilde{\vec{v}} \in S \cap \complement{N''}} \\
    &\le \pr{\vec{v} + \widetilde{\vec{v}} \in N''} + \pr{\vec{v} + \widetilde{\vec{v}} \in S \cap \complement{N''}} \\
    &\le \pr{\vec{v} + \widetilde{\vec{v}} \in N''} + \pr{\vec{v} + \widetilde{\vec{v}} \not\in N''} \\
    &\le e^\epsilon \pr{\vec{v}' + \widetilde{\vec{v}} \in S} + \delta \\
\end{align*}

\section{Proof of Theorem~\ref{thm:differential_privacy_general}}
\label{appendix:proof_dp_general}
To see how many borrowers would be affected by a specific borrower $b$, we first arrange the dataset consisting of $m$ elements as a $m_1 \times m_2 \times \dots \times m_d$ array: $A = \{a_{k_1k_2\dots k_d; 1 \le k_i \le m_i}\}$.
Now we consider two borrowers $b = a_{k_1k_2\dots k_d}$ and $b' = a_{k_1'k_2'\dots k_d'}$. 

We first assume that after the $(s-1)$-th iteration of a PIR query, we get $\hat{m}_{s-1} = m/\prod_{i=1}^{s-1}m_i$ ciphertexts, say $C^{s-1} = \{c_{k_{s}k_{s+1}\dots k_d}^{s-1}; 1 \le k_i \le m_i\}$, and some of them may be $0$'s, i.e. empty items.
Then we perform the $s$-th iteration using the subquery ${q_{s1}, q_{s2}, \dots, q_{sm_{s}}}$ and get $C^s = \{c_{k_{s+1}k_{s+2}\dots k_d}^{s}; 1 \le k_i \le m_i\}$ consisting of $\hat{m}_{s}=\hat{m}_{s-1}/m_{s}$ elements, 
where $c_{k_{s+1}k_{s+2}\dots k_d}^{s} = \prod_{1 \le j \le k_{s},c_{jk_{s+1}\dots k_d}^s \ne 0}{(c_{jk_{s+1}\dots k_d}^{s-1})^{q_{sj}}}$.
This indicates that the sequence of ciphertexts involving $a_{k_1k_2\dots k_d}$ is $S = (c_{k_1k_2\dots k_d}^1, c_{k_2\dots k_d}^2, \dots, c_{k_d}^{d-1}, c^d)$, where $c^d$ is the final PIR response exposed to the receiver.
It can be seen that $c_{k_{s}k_{s+1}\dots k_d}^{s-1}$ is only involved in $c_{k_{s+1}k_{s+2}\dots k_d}^{s}$, that is, a ciphertext in $C^{s-1}$ is only involved in one ciphertext in $C^s$.
We can then use mathematical induction to prove that an element in $A$ can only be involved in one ciphertext in the $s$-th iteration ciphertexts $C^s$.

Now we consider the situation where the query is for $b = a_{k_1k_2\dots k_d}$, and we want to know how $b' = a_{k_1'k_2'\dots k_d'}$ affects the query response. 
The ciphertext sequences involving $b$ and $b'$ are $S = (c_{k_2k_3\dots k_d}^1, c_{k_3\dots k_d}^2, \dots, c_{k_d}^{d-1}, c^d)$ and $S' = (c_{k_2'k_3'\dots k_d'}^1, c_{k_3'\dots k_d'}^2, \dots, c_{k_d'}^{d-1}, c^d)$, repectively .

We first find a minimum $i$ such that $k_{i+1} = k_{i+1}', k_{i+2} = k_{i+2}', \dots, k_{d} = k_{d}'$.
If such $i$ exists, we have $c_{k_{i}k_{i+1}\dots k_d}^{i-1} \ne c_{k_{i}'k_{i+1}'\dots k_d'}^{i-1}$ and $c_{k_{i+1}k_{i+2}\dots k_d}^i = c_{k_{i+1}'k_{i+2}'\dots k_d'}^i$.
According to the above analysis, how $b'$ affects $c^d$ is equivalent to how $c_{k_{i}'k_{i+1}'\dots k_d'}^{i-1}$ affects $c_{k_{i+1}k_{i+2}\dots k_d}^{i}$.
As the query is for $b$, the query ciphertext for $c_{k_{i}'k_{i+1}'\dots k_d'}^{i-1}$ is $E(0)$, while the query ciphertext for $c_{k_{i}k_{i+1}\dots k_d}^{i-1}$ is $E(1)$.
Then there are three cases:

\begin{enumerate}[leftmargin=0.5cm, label=\arabic*)]
\item $c_{k_{i}k_{i+1}\dots k_d}^{i-1} \ne 0$, i.e. $c_{k_{i}k_{i+1}\dots k_d}^{i-1}$ is not empty: in this case, we always have $c_{k_{i+1}k_{i+2}\dots k_d}^i = E(c_{k_{i}k_{i+1}\dots k_d}^{i-1})$.
Thus $c_{k_{i}'k_{i+1}'\dots k_d'}^{i-1}$ has no effect on $c_{k_{i+1}k_{i+2}\dots k_d}^i$.

\item $c_{k_{i}k_{i+1}\dots k_d}^{i-1}$ is empty but there exists a non-empty item in the set $C^s \setminus \{c_{k_{i}k_{i+1}\dots k_d}^{i-1}, c_{k_{i}'k_{i+1}'\dots k_d'}^{i-1}\}$: 
In this case, we always have $c_{k_{i+1}k_{i+2}\dots k_d}^i = E(0)$, thus $c_{k_{i}'k_{i+1}'\dots k_d'}^{i-1}$ has no effect on $c_{k_{i+1}k_{i+2}\dots k_d}^i$.

\item $c_{k_{i}k_{i+1}\dots k_d}^{i-1}$ as well as all the elements in the set $C^s \setminus \{c_{k_{i}k_{i+1}\dots k_d}^{i-1}, c_{k_{i}'k_{i+1}'\dots k_d'}^{i-1}\}$ are empty: 
In this case, if $c_{k_{i}'k_{i+1}'\dots k_d'}^{i-1}$ is empty, $c_{k_{i+1}k_{i+2}\dots k_d}^i = 0$, otherwise $c_{k_{i+1}k_{i+2}\dots k_d}^i = E(0)$. 
This means $c_{k_{i}'k_{i+1}'\dots k_d'}^{i-1}$ has an effect on $c_{k_{i+1}k_{i+2}\dots k_d}^i$ in this case.
\end{enumerate}

But if there is no such $i$ that $k_{i+1} = k_{i+1}', k_{i+2} = k_{i+2}', \dots, k_{d} = k_{d}'$, the only way $b'$ can affect the value of $c^d$ is to make $c_{k_d'}^{d-1}$ affect $c^d$. However, as we always replace empty result $0$ with $E(0_{d-1})$, $b'$ cannot affect the value of $c^d$.

In conclusion, $b$ affects $c^d$ if and only if there is a such $i$ and at the same time the absence of $c_{k_{i}'k_{i+1}'\dots k_d'}^{i-1}$ would make $c_{k_{i+1}k_{i+2}\dots k_d}^i = 0$ empty.
It is easy to see that the existence of such $i$ is equivalent to $k_d = k_d'$. 
For a specific borrower $b = a_{k_1k_2\dots k_d}$, the number of borrowers meeting this requirement is $\prod_{i=1}^{d-1}m_i$.
So we conclude a) in Theorem~\ref{thm:differential_privacy_general}.

On the other hand, replacing the empty items with $E^{d-s}(0_s)$ after the $s$-th iteration, the absence of $c_{k_{s}'k_{s+1}'\dots k_d'}^{s-1}$ would have no effect on $c_{k_{s+1}k_{s+2}\dots k_d}^{s}$ and its successors any more, as case 3) above would not appear after the $s$-th iteration.
In this situation, to affect $c^d$, $b'$ should affect at least one of the elements in $(c_{k_2k_3\dots k_d}^1, c_{k_3\dots k_d}^2, \dots, c_{k_sk_{s+1}\dots k_d}^{s-1})$, which is equivalent to $k_s = k_s', k_{s+1}=k_{s+1}', \dots, k_d = k_d'$.
For a specific borrower $b = a_{k_1k_2\dots k_d}$, the number of borrowers meeting the requirement is $\prod_{i=1}^{s-1}m_i$.
Meanwhile, as all the empty items are replaced after the $s$-th iteration, there would be no empty items since then. In other words, th final output $c^d$ would not be of $type\ {s+1}, type\ {s+2}, \dots, type\ d$. Thus we have $\hat{d} = s$.
So we conclude b) in Theorem~\ref{thm:differential_privacy_general}.

\section{Secure Aggregation Protocol}
\label{appendix:pir_sparse_recursive}
Protocol~\ref{protocol:pir_sparse_recursive} summarizes our secure aggregation protocol.
 The correctness of the consistency check in this protocol is directly from \cite{ZKP_HOMO}. Also, introducing the noise from the exchanger does not hurt the privacy, as we can treat the exchanger as another lender which generates responses containing commitments to $0$. That is, as long as the originator correctly retrieves the commitments from the lenders, the borrower is unable to cheat the originator about her loan amount.
 Meanwhile, the randomness added to each ciphertext in Algorithm~\ref{algorithm:pir_sparse_recursive} and the noise responses generated by the exchanger prevent the originator from inferring extra information about each lender's data from the responses.

\begin{protocol}[htbp]
    \DontPrintSemicolon
    \begin{enumerate}[label={\alph*)}, leftmargin=0.1cm, rightmargin=0.3cm, itemsep=0mm]
        \item To retrieve the loan information of a borrower $b$, the originator generates a $d$-dimentional recursive query and sends it along with $gid_b$ to the lenders through the exchanger.
        \item For each registered user $u$ who is in the group $gid_b$ and has borrowed money $x_{iu}$ from the lender $S_i$, $S_i$ calculates $r_{iu} = \text{PRF}_{\tau_{iu}}(``rc" \| u \| x_{iu} \| \text{date})$ and generates the commitment to $x_{iu}$ as $c_{iu} = F(x_{iu}, r_{iu})$.  Then it arranges the commitments in an array $A$ according their $pid$'s, and performs the computation in Algorithm~\ref{algorithm:pir_sparse_recursive}.
        \item Each lender sends its result to the exchanger: if Algorithm~\ref{algorithm:pir_sparse_recursive} returns a ciphertext, it sends the ciphertext to the exchanger; otherwise, if Algorithm~\ref{algorithm:pir_sparse_recursive} returns $0$, it sends $\text{E}^d(0)$ to the exchanger.
        \item For each $i = 1, 2, \dots, d$, the exchanger generates $\widetilde{n_i}$ noise responses of \type{i}, where $\widetilde{n_i} \sim \lceil \max(0, \lap{\mu, \lambda}) \rceil$. 
            Meanwhile, the exchanger samples a random integer $n_0 \sim \lceil \max(0, \lap{\mu, \lambda}) \rceil$, and genereates $n_0$ commitments $F(0, r_1), F(0, r_2), \dots, F(0, r_{n_0})$. Then the exchanger calculates $r_z = \sum_{j = 0}^{n_0} r_j$, and encrypts the commitments to get $n_0$ responses of \type{0}. Finally, the exchanger mixes these responses with the responses collected from the lenders.
        \item The borrower calculates $r_o = \text{PRF}_{\tau_{ob}}(``rc" \| b \| \text{date})$.
              For each $S_i$, the borrower $b$ calcluates $r_i = \text{PRF}_{\tau_{ib}}(``rc" \| b \| x_{ib} \| \text{date})$. 
              Also, the borrower generates a commitment to the total money she has borrowed: $c_b = F(\sum x_{ib}, r_b)$. Then the borrower sends $\Delta r_b = r_b - r_o - \sum r_i$ to the exchanger and sends $c_b$ to the originator through the exchanger.
        \item The exchanger shuffles the responses and sends the responses to the originator.
              Also, the exchanger calculates $\Delta r = \Delta r_b - r_z$, then sends $\Delta r$ to the originator.
        \item The originator initializes an empty set $C$ and decrypts the responses. If a response is of $type \ 0$, the originator adds the commitment contained in the response to $C$. Then the originator calculates $c = \prod_{c_i \in C}c_i$. Finally, the originator checks if $c_b = c\cdot h^{\Delta r + r_o}$, where $r_o = \text{PRF}_{\tau_{ob}}(``rc" \| b \| \text{date})$.
    \end{enumerate}
    \caption{Secure aggregation using DSPIR, $\Pi_{agg}$}
    \label{protocol:pir_sparse_recursive}
\end{protocol}

\section{ZKP for a Valid Query}
\label{appendix:valid_query}

Algorithm~\ref{algorithm:zkp_right_form} shows the zero-knowledge proof of a valid query. 
The goal is that, for the subquery $q_{i1}, q_{i2}, \dots, q_{im_i}$ of each dimension $i$, we prove that there is only one $1$ in the subquery and the others are $0$'s.
We first prove that each $q_{ij}$ encrypts $0$ or $1$ using partial ZKP, then prove that the sum of the subquery is $1$, i.e., $\prod_jq_{ij}$ encrypts $1$.
This indicates that there is only one $1$ in each subquery. \\

\begin{algorithm}[htbp]
    \DontPrintSemicolon
    \begin{itemize}[label={}, leftmargin=0.0cm, itemsep=0mm, topsep=0mm] \item \textbf{Initialize:} \end{itemize}
    The prover and the verifier share a recursive PIR query $q = \{\{q_{11}, q_{12}, \dots, q_{1m_1}\}, \{q_{21}, q_{22}, \dots, q_{2m_2}\}, \dots, $ \\ $ \{q_{d1}, q_{d2}, \dots, q_{dm_d}\}\}$.
    \begin{itemize}[label={}, leftmargin=0cm, itemsep=0mm, topsep=0mm] \item \textbf{ } \end{itemize}

    \begin{itemize}[label={}, leftmargin=0.0cm, itemsep=0mm, topsep=0mm] \item \textbf{Prover:} \end{itemize}
    \begin{enumerate}[label={\alph*)}, leftmargin=0.1cm, rightmargin=0.1cm, itemsep=0mm, topsep=0mm]
        \item For each $i \in \{1, 2, \dots, d\}$, calculate $q_i = \prod_{j=1}^{m_j} q_{ij}$.
        \item For each $i \in \{1, 2, \dots, d\}$ and each $j \in \{1, 2, \dots, m_i\}$, generate non-interactive partial ZKP $p_{ij}$ which proves \\ that $q_{ij}$ contains either $0$ or $1$, and non-interactive ZKP \\ $p_i$ which proves that $q_i$ contains $1$.
        \item Send $p = \{\{p_{11}, p_{12}, \dots, p_{1m_1}, p_1\}, \{p_{21}, p_{22}, \dots, p_{2m_2}, p_2\}, $ \\ $ \dots, \{p_{d1}, p_{d2}, \dots, p_{dm_d}, p_d\}\}$ as the proof to the verifier.
    \end{enumerate}
    \begin{itemize}[label={}, leftmargin=0cm, itemsep=0mm, topsep=0mm] \item \textbf{ } \end{itemize}

    \begin{itemize}[label={}, leftmargin=0cm, itemsep=0mm, topsep=0mm] \item \textbf{Verifier:} \end{itemize}
    \begin{enumerate}[label={\alph*)}, leftmargin=0.1cm, rightmargin=0.1cm, itemsep=0mm, topsep=0mm]
        \item For each $i \in \{1, 2, \dots, d\}$ and each $j \in \{1, 2, \dots, m_i\}$, \\ use $p_{ij}$ to verify that $q_{ij}$ indeed contains $0$ or $1$.
        \item For each $i \in \{1, 2, \dots, d\}$, calculate $q_i = \prod_{j=1}^{m_j} q_{ij}$, \\ and use $p_i$ to verify that $q_i$ indeed contains $1$.
        \item Return true if the above verifications pass.
    \end{enumerate}
    \caption{Zero-knowledge proof of a valid query, $\Pi_{ZKPoQ}$.}
    \label{algorithm:zkp_right_form}
\end{algorithm}

\noindent \textsc{Proof Sketch: }
    Our zero-knowledge proof of a valid query is based on common ZKP techniques (i.e. partial ZKP and ZKP for Paillier cryptosystem), thus the completeness and zero-knowledgeness of our proof comes directly from these techniques.

    And we show the soundness as follows. First we can see that each $q_{ij}$ encrypts $0$ or $1$, which is proved using partial ZKP. 
    We then assume that the subquery of each dimension $i$, $q_{i1}, q_{i2}, \dots, q_{im_i}$, contains $k_i$ $1$'s and $m_i - k_i$ $0$'s. Thus for a valid query $q$, we have $k_i = 1$ for every $i$.
    In other words, if the query $q$ is invalid, there should exist a dimension $i$ such that $k_i \ne 1$.
    However, for each dimension $i$, $q_i = \prod_{j=1}^{m_j} q_{ij}$ encrypts the sum of the number encrypted in $q_{ij}$, and $\prod_{j=1}^{m_j} q_{ij}$ actually encrypts $k_i$ in our situation. Thus if $k_i \ne 1$, the prover cannot prove that $q_i$ encrypts $1$. This means that only when all the subqueries are valid an the prover prove that $q_i$'s encrypt $1$.
$\qed$

\section{ZKP of the Correspondence of a Secret and a Recursive PIR Query}
\label{appendix:correspondence}
The prover sends a $d$-dimensional recursive PIR query $q$ and a ciphertext $c$ to the verifier. Then the verifier performs PIR on an array $A$ consisting of $m$ items using the query $q$ and finally get $\text{E}^d(a)$, where $a$ is an item of $A$.
The goal of the prover in this protocol is to prove that the number $a$ encrypted in $\text{E}^d(a)$ is also encrypted in $c$. Algorithm~\ref{algorithm:zkp_secret_recursive_PIR} shows this process. 
We denote $K$ as the index set $\{(k_1, k_2, \dots, k_d)\}$ where $1 \le k_i \le m_i$. And for each $i \in \{1, 2, \dots, d\}$, we denote $K^*_i$ as the set $\{(k_1, \dots, k_{i-1}, k_{i+1}, \dots, k_d)\}$. Obliviously, $|K| = m$ and $|K^*_i| = m / m_i$. Also, for a $k^* = (k_1, \dots, k_{i-1}, k_{i+1}, \dots, k_d) \in K^*_i$, we denote $a_{k_ik^*} = a_{k_1 \dots k_{i-1} k_i k_{i+1} \dots k_d}$, an element of $A$ with index $(k_1, \dots, k_{i-1}, k_i, k_{i+1}, \dots, k_d)$.

We need a protocol to prove that one of the given several ciphertexts $c_1, c_2, \dots, c_m$ encrypts the same number as another ciphertext $c$, i.e., $\exists i \in \{1, 2, \dots, m\}$, the plaintext of $c_i$ is the same as the plaintext of $c$.
This can be done using existing techniques. We first consider a simple situation where $m=1$ and the prover needs to prove that $c_1$ encrypts the same number as $c$. Given $c_1 = g^xr_1^n$ and $c = g^xr^n$, the prover can prove that $c_1$ encrypts the same number as $c$ by showing that she knows a number $r'$ which is a $n$-th modulo root of $c_1c^{-1}$ (equivalent to $(r_1r^{-1})^n$) using the techniques in \cite{ZKP_HOMO, PAILLIER_ZKP}. 
It can be seen that, to accomplish this proof, the prover only needs to know $r_1r^{-1}$, and does not need to know $c_1, c$ or $x$. 
Combining this proof and the partial ZKP technique~\cite{PARTIAL_ZKP}, we can get the target proof.
\\

\begin{algorithm}[htbp]
    \DontPrintSemicolon
    \begin{itemize}[label={}, leftmargin=0.0cm, itemsep=0mm, topsep=0mm] \item \textbf{Initialize:} \end{itemize}
    The prover and the verifier share a valid recursive PIR query $q = \{\{q_{11}, q_{12}, \dots, q_{1m_1}\}, \{q_{21}, q_{22}, \dots, q_{2m_2}\}, \dots, $ \\ $ \{q_{d1}, q_{d2}, \dots, q_{dm_d}\}\}$, an array $A$ with $m$ different items, and a ciphertext $c$ which is the ciphertext of one of the items of $A$. 
    The prover knows the plaintext $x_{ij}$ and the randomness $r_{ij}$ of each $q_{ij}$ (i.e. $q_{ij} = g^{x_{ij}} r_{ij}^n$), as well as the randomness $r$ of $c$, while the verifier does not know these information.
    \begin{itemize}[label={}, leftmargin=0cm, itemsep=0mm, topsep=0mm] \item \textbf{ } \end{itemize}

    \begin{itemize}[label={}, leftmargin=0.0cm, itemsep=0mm, topsep=0mm] \item \textbf{Prover:} \end{itemize}
    \begin{enumerate}[label={\alph*)}, leftmargin=0.1cm, rightmargin=0.1cm, itemsep=0mm, topsep=0mm]
        \item Arrange $A$ as a $m_1 \times m_2 \times \dots \times m_d$ array: $A = \{a_{k_1 k_2 \dots k_d}\}$ where $1 \le k_i \le m_i$ for each dimension $i$.
        \item Generate a list of random numbers $\{\{r_{11}, r_{12}, \dots, r_{1m_1}\}, $ \\ $ \{r_{21}, r_{22}, \dots, r_{2m_2}\}, \dots, \{r_{d1}, r_{d2}, \dots, r_{dm_d}\}\}$. \\  
            For each dimension $i (1 \le i \le d)$ and each $k^*_{ij} \in K^*_i (1 \le j \le |K^*_i|)$, calculate $r^*_{ij} = \prod_{k_i=1}^{m_i} ({r_{ik_i}})^{a_{k_ik^*_{ij}}}$.
        \item For each dimension $i$, using $r^*_{i1}, r^*_{i2}, \dots, r^*_{i|K^*_i|}$ and $r$, generate the proof which proves that one of $c^*_{i1}, c^*_{i2}, \dots, c^*_{i|K^*_i|}$ encrypts the same number as $c$.
            \\ Send the proof to the verifier.
    \end{enumerate}
    \begin{itemize}[label={}, leftmargin=0cm, itemsep=0mm, topsep=0mm] \item \textbf{ } \end{itemize}

    \begin{itemize}[label={}, leftmargin=0cm, itemsep=0mm, topsep=0mm] \item \textbf{Verifier:} \end{itemize}
    \begin{enumerate}[label={\alph*)}, leftmargin=0.1cm, rightmargin=0.1cm, itemsep=0mm, topsep=0mm]
        \item Arrange $A$ as a $m_1 \times m_2 \times \dots \times m_d$ array: $A = \{a_{k_1 k_2 \dots k_d}\}$.
        \item For each dimension $i (1 \le i \le d)$, use the subquery $\{q_{i1}, q_{i2}, \dots, q_{im_i}\}$ to perform information retrieval along that dimension, which outputs $m / m_i$ ciphertexts: for each $k^*_{ij} \in K^*_i$, calculate $c^*_{ij} = \prod_{k_i=1}^{m_i} ({q_{ik_i}})^{a_{k_ik^*_{ij}}}$, equivalent to $\prod_{k_i=1}^{m_i} ({g^{x_{ik_i}}r_{ik_i}^n})^{a_{k_ik^*_{ij}}} = g^{\sum_{k_i=1}^{m_i} x_{ik_i}a_{k_ik^*}}{r^*_{ij}}^n$.
        \item For each dimension $i$, using the proof from the prover, verify that one of $c^*_{i1}, c^*_{i2}, \dots, c^*_{i|K^*_i|}$ encrypts the same number as $c$.
    \end{enumerate}
    \caption{Zero-knowledge proof of the correspondence of a secret and a recursive PIR query, $\Pi_{ZKPoCR}$}
    \label{algorithm:zkp_secret_recursive_PIR}
\end{algorithm}

\noindent \textsc{Proof Sketch: }
    The zero-knowledgeness of this proof comes directly from the partial ZKP technique. 
    As for the completeness, as the query is valid, let we assume that, for each dimension $i$, we have $q_{ik_i'} = \enc{1}$, i.e., $x_{ik_i'} = 1$. Thus, $c^*_{ij} = g^{\sum_{k_i=1}^{m_i} x_{ik_i}a_{k_ik^*}}{r^*_{ij}}^n = g^{a_{k'_ik^*}}{r^*_{ij}}^n = \enc{a_{k'_ik^*}}$. That is, for each dimension $i$, $c^*_{i1}, c^*_{i2}, \dots, c^*_{i|K^*_i|}$ are the ciphertexts of all the items with $k_i = k_i'$. And if $c$ really encrypts the number corresponding to the query, we have $c = \enc{a_{k_1'k_2' \dots k_d'}}$. Thus the number encrypted in $c$ must in the set  $\{ c^*_{i1}, c^*_{i2}, \dots, c^*_{i|K^*_i|} \}$, i.e., $c$ encrypts the same number as one of $c^*_{i1}, c^*_{i2}, \dots, c^*_{i|K^*_i|}$.

    We then show the soundness as follows. Let we assume that the number $c$ encrypts is $a_{k_1''k_2''\dots k_d''}$. Now if there is a dimension $i$ such that $k_i'' \ne k_i'$, then the proof that one of $c^*_{i1}, c^*_{i2}, \dots, c^*_{i|K^*_i|}$ encrypts the same number as $c$ would fail, as $c^*_{i1}, c^*_{i2}, \dots, c^*_{i|K^*_i|}$ are the ciphertexts of the items with $k_i = k_i'$ and the items in $A$ are of different values. Thus, if all the proofs pass, we have $k_i'' = k_i'$ for every dimension $i$, which indicates that $a_{k_1''k_2''\dots k_d''} = a_{k_1'k_2'\dots k_d'}$, i.e., the number encrypted in $c$ corresponds to the query.
$\qed$

\section{Anonymous Authorization Protocol}
\label{appendix:anonymous_authorization}
Protocol~\ref{protocol:anonymous_authorization} shows the overall protocol for anonymous authorization. 
The borrower and the originator use a PRF to generate randomness for Paillier encryption. The PRF takes $\tau_{ob}$ shared between the borrower and the originator as seed.
Note that instead of using $\tau_{eu}$'s directly as the secrets to be retrieved, the exchanger generates a new random number $r_e$ in each round and uses another PRF to output random numbers $y_u$'s as the secrets to be retrieved in this round. As the PRF takes $\tau_{eu}$'s as seeds, each registered user $u$ shares a distinct $y_u$ with the exchanger. 
The reason why we do not use $\tau_{eu}$'s directly is that the ZKP protocol for correspondence mentioned above requires the verifier (i.e. the exchanger) to send the secrets to the prover (i.e. originator). Thus, we should use newly generated random numbers as the secrets to be retrieved by the PIR query in each round to avoid the leakage of $\tau_{eu}$'s.

\begin{protocol}[htbp]
    \DontPrintSemicolon
    \begin{enumerate}[label={\alph*)}, leftmargin=0.1cm, rightmargin=0.3cm, itemsep=0mm]
        \item The exchanger generates a random number $r_e$ and sends $r_e$ to the borrower.
        \item Unpon receiving $r_e$, the borrower $b$ calculates $y = \text{PRF}_{\tau_{eb}}(``y" \| r_e \| \text{date})$.
            With the Paillier public key $pk_o = (n, g)$ of the originator , the borrower encrypts $y$ as $c = g^yr^n$, where $r = \text{PRF}_{\tau_{ob}}(``r" \| pk_o \| b \| \text{date})$. 
            Then the borrower sends $c$ and the zero-knowledge proof $P_k$ which proves to the exchanger that she knows the plaintext (i.e. $y$) in $c$.
        \item For each user $u$ in the group $gid_b$, the exchanger calculates $y_u = \text{PRF}_{\tau_{eu}}(``y" \| r_e \| \text{date})$ and sends $y_u$'s to the originator.
        \item The originator generates the PIR query $q$
            and sends the query and the zero-knowledge proof $P_q$ which proves that the query is valid to the exchanger.  
        \item The originator calculates $r = \text{PRF}_{\tau_{ob}}(``r" \| pk \| b \| \text{date})$. With $q$, $r$ and $y_u$'s, the originator generates the zero-knowledge proof $P_{cr}$, which proves that for each $i \in \{1, 2, \dots, d\}$, the ciphertext $c$ encrypts the same number as one of ciphertexts produced by the subquery of dimension $i$. Then the originator sends the proof to the exchanger.
        \item Finally, with the proofs from the borrower and the originator, the exchanger verifies three things:
            1) The borrower really knows the plaintext encrypted in $c$;
            2) The query from the originator is valid;
            3) For each $i \in \{1, 2, \dots, d\}$, the ciphertext $c$ encrypts the same number as one of ciphertexts produced by the subquery of dimension $i$.
    \end{enumerate}
    \caption{Anonymous borrower authorization, $\Pi_{auth}$}
    \label{protocol:anonymous_authorization}
\end{protocol}

\section{Security Analysis and Proof of \sysname}
\label{appendix:formal_proof}

For security assumption, we assume that the exchanger and lenders are semi-honest, while assuming that the borrower and originator are malicious, as what is stated in Section~\ref{section:threat_model}. In the following proofs, we denote this assumption about participants by $\mathcal{P}_{\sysname}$.
In addition, we use $ex$, $og$, $bo$ and $le$ as the abbreviations of exchanger, originator, borrower and lender, respectively.
For each protocol, we prove three properties: 
1) \textbf{correctness}: the protocol gives expected outputs and detects malicious behaviors;
2) \textbf{privacy}: the output reveals no unexpected information;
3) \textbf{security}: during the execution of the protocol, an participant gets no information other than the output.
Specifically, we define the security of the protocols in this paper as follows:

\begin{defi} \label{def:security}
A protocol $\pi$ securely realizes a function $f$ in the presence of $\mathcal{P}_{\sysname}$, 
    if for every probabilistic polynomial-time adversary $i \in \{ex, og, bo, le\}$, there exists a probabilistic polynomial-time simulator $\mathcal{S}$ such that for every possible input tuple $\vec{x}$, we have:
\begin{center}
    $\mathcal{S}(x_i, f_i(\vec{x})) \approx \text{View}_{i}^{\pi}(\vec{x})$
\end{center}
    where $f_i(\vec{x})$ is the output for $i$ of $f$.
\end{defi}

We prove the security with the hybrid model by showing that our protocols satisfy Definition~\ref{def:security}.
We first prove that Protocol~\ref{protocol:anonymous_authorization} $\Pi_{auth}$ is secure.
Then we prove the security of Protocol~\ref{protocol:pir_sparse_recursive} $\Pi_{agg}$ when composed with $\Pi_{auth}$.
Finally, we show that the security of our protocol satisfies the security requirements in Section~\ref{section:security_goal}.

\subsection{Security Proof of Anonymous Authorization}

We prove the security of $\Pi_{auth}$ in this section. 
We define $\mathcal{F}_{auth}$ as the functionality of $\Pi_{auth}$ as follows. 
The output indicates whether the query the originator sends is authorized.

\noindent \begin{minipage}[h]{0.47\textwidth}
\fbox{
\parbox{\textwidth}{
\begin{center}
Functionality $\mathcal{F}_{auth}$
\end{center}

\para{Inputs: }
The borrower inputs the secret $\tau_{eb}$.
The originator inputs a $d$-dimensional plaintext query $q$ and the group id $gid_b$ of the borrower.
The exchanger inputs a set of the secrets $T_E = \{\tau_{eu}\}$ shared between the exchanger and the users/borrowers.

\para{Process: }
\begin{enumerate}[leftmargin=0.5cm, label=\arabic*.]
\item Check that $q$ is valid. If the check fails, set $z = \bot$, output $z$ to the exchanger and abort.
\item Retrieve $\tau$ from $T_E$ according to $gid_b$ and $q$.
\item Check that the retrieved value $\tau$ equals $\tau_{eb}$. If the check succeeds, set $z = ``pass"$, otherwise set $z = \bot$.
\item Output $z$ to the exchanger.
\end{enumerate}
}
}
\end{minipage}

\begin{thm}
\label{thm:auth_security}
    $\Pi_{auth}$ securely realizes $\mathcal{F}_{auth}$ in the presence of $\mathcal{P}_{\sysname}$.
\end{thm}

\noindent \textsc{Proof Sketch: }
\para{Correctness. }
The joint input of the originator and the borrower in $\Pi_{auth}$ includes a ciphertext $c$, a PIR query $q$ and proofs $P_k$, $P_q$ and $P_{cr}$ (we sometimes omit $gid_b$ as it does not affect the correctness of our analysis). The exchanger verifies the three proofs in step $f$ of $\Pi_{auth}$. 
Specifically, the exchanger uses $c$ and $P_k$ to verify proof 1, uses $q$ and $P_q$ to verify proof 2, and uses $c$, $q$ and $P_{cr}$ to verify proof 3.
With the ZKP protocols describted in the above sections, honest inputs enable the verifications to pass, while any dishonest value in the set $\{c, q, P_k, P_q, P_{cr}\}$ causes verification failure.
Specifically, proof 2 ensures a valid query, while proof 1 and 3 ensure that the borrower knows the plaintext in the ciphertext retrieved using the query, thus eradicate a malicious originator and borrower who does not know $\tau_{eb}$.

\para{Privacy. }
The output is an indicator indicating that the query is authorized by a real borrowr or not. The indicator itself reveals no sensitive information about the participants' private input.

\para{Security. }
We first prove the security against a malicious originator and a malicious borrower.
For an adversary $\mathcal{A}_o$ that corrupts the originator,
the view of $\mathcal{A}_o$ in $\Pi_{auth}$ is a set of pseudo-random numbers $Y = \{y_u\}$.
We then construct a simulator $\mathcal{S}_o$ that produces a set of random numbers $Y' = \{y_u'\}$.
As each pseudo-random number $y_u \in Y$ in $\Pi_{auth}$ is generated using a seed that is not known by the originator, 
when the pseudo-random function $\text{PRF}$ we use is secure, $\mathcal{A}_o$ can only distinguish $Y$ and $Y'$ with negligible probability.

For an adversary $\mathcal{A}_b$ that corrupts the borrower,
the view of $\mathcal{A}_b$ in $\Pi_{auth}$ is a random number $r_e$.
We can construct a simulator $\mathcal{S}_b$ that produces a random number $r_e'$.
The indistinguishability is direct.

However, if an adversary $\mathcal{A}_{ob}$ corrupts both the originator and the borrower, the joint view of $\mathcal{A}_{ob}$ includes $Y$ and $r_e$.
We then construct a probabilistic-polynomial time simulator $\mathcal{S}_{ob}$ which works as follows:
\begin{enumerate}[leftmargin=0.5cm, label=\arabic*.]
\item Generate a pesudo-number $y_b' = \text{PRF}_{\tau_{eb}}(``y" \| r_e \| \text{date})$. For every other user $u$ in group $gid_b$, generate a random number $y_u$.
     Compose these numbers into a set $Y'$.
\item Generate a random number $r_e'$.
\item Send $Y'$ and $r_e'$ to the adversary $\mathcal{A}_{ob}$.
\end{enumerate}
Now we show $(Y, r_e) \approx (Y', r_e')$. With the knowledge of $\tau_{eb}$, $\mathcal{A}_{ob}$ can generate $y_b$ such that $y_b = y_b'$.
The other pseudo-random numbers in $Y$, however, are generated using seeds not known by $\mathcal{A}_{ob}$, and thus are computationally indisinguishable with the truely-random numbers in $Y'$ for $\mathcal{A}_{ob}$.
Meanwhile, both $r_e$ and $r_e'$ are truely-random numbers, and thus are indisinguishable. 
This indicates the indistinguishability.

For the exchanger, the view is a tuple $(c, q, P_k, P_q, P_{cr})$. We construct a probabilistic-polynomial time simulator $\mathcal{S}_e$ that receives $gid_b$, $T_E$ and $z$ from the exchanger.
And we consider two cases:
a) $z$ is $\bot$; b) $z$ is $``pass"$.

Case $a$ indicates that the borrower $b$ intends to pretend another borrower $b'$.
In this situation, the borrower and the originator do not have the private information of $b'$ (i.e. the secret $\tau_{eb'}$ shared between $b'$ and the exchanger) and the security analysis for this case is trival: $\mathcal{S}_e$ just needs to behave as the same as the borrower $b$ and originator.
We omit the details.

For case $b$, both the borrower and originator behave honestly. $\mathcal{S}_e$ works as follows:
\begin{enumerate}[leftmargin=0.5cm, label=\arabic*.]
    \item For each user $u$ in group $gid_b$, generate a pesudo-number $y_u' = \text{PRF}_{\tau_{eb}}(``y" \| r_e \| \text{date})$, and denote $Y' = \{y_u'\}$.
          Randomly choose a position $u'$ in that group and select a corresponding number $y'$ in $Y'$. 
      \item Encrypt $y'$ as $c' = g^{y'}{r'}^n$, where $(n, g)$ is a random public key of Paillier crypto system and $r'$ is a random number sampled from $\mathbb{Z}_n$. Then generate a proof $P_k'$ that proves the knowledge of the plaintext encrypted in $c'$. Output $c'$ and $P_k'$ to the exchanger.
    \item Generate a $d$-dimensional PIR query $q'$ according to $u'$. Generate a proof $P_q'$ that proves $q'$ is a valid query. Output $q'$ and $P_q'$ to the exchanger.
    \item Use $q'$, $r'$ and $Y'$ to generate a proof $P_{cr}'$ that proves the correspondence. Output $P_{cr}'$ to the exchanger.
\end{enumerate}
We need to prove that $(c, q, P_k, P_q, P_{cr}) \approx (c', q', P_k', P_q', P_{cr}')$.
We can treat $(c, P_k)$ as the output of a randomized function $f$ which takes $y$ as input, i.e. $(c, P_k) = f(y)$ and $(c', P_k') = f(y')$.
We first see that $c \approx c'$ due to the security of the Paillier crypto system. 
If an adversary can distinguish between $(c, P_k)$ and $(c', P_k')$ with advantage $\epsilon$, he can distinguish which input ($y$ or $y'$) $f$ takes with advantage $\epsilon$.
However, according to the \emph{zero knowledge} property of $\Pi_{ZKPoK}$, $\epsilon$ should be negligible.
The same applies to proving $(q, P_q) \approx (q', P_q')$ and $(c, q, P_{cr}) \approx (c', q', P_{cr}')$.
It remains to show that composing these proofs retains indistinguishability.
Actually, each private number behind each proof is masked by a random number, and the random number is hiden using a one-way function that cannot be opened by the verifier (please see \cite{ZKP_HOMO} for details). 
As in our protocol these random numbers are generated independently, the three proofs are independent random tuples for the exchanger, thus each proof would not hurt the zero knowledge property of the others. 
Finally, we have $(c, q, P_k, P_q, P_{cr}) \approx (c', q', P_k', P_q', P_{cr}')$.
$\qed$

\subsection{Security Proof of Secure Aggregation}

\noindent \begin{minipage}[h]{0.47\textwidth}
\fbox{
\parbox{\textwidth}{
\begin{center}
Functionality $\mathcal{F}_{agg}$
\end{center}
\para{Initialization: }
Upon invocation, $\mathcal{F}_{agg}$ gets the random tape $R_e$ of the exchanger and the common input $N_g$.

\para{Inputs: }
The borrower inputs a random number $r_b$ and a commitment $c_b$.
The originator inputs a $d$-dimensional plaintext query $q$ and the group id of the borrower $gid_b$.
Each lender $S_i$ inputs its database $D_i$.

\para{Process: }
\begin{enumerate}[leftmargin=0.5cm, label=\arabic*.]
\item Initialize a type count vector $\vec{v} = \{n_0, n_1, \dots, n_d\}$ where for each $0 \le j \le d$ we have $n_j = 0$.
      For each $D_i$, deduce the type of ciphertext by simulating the PIR process using the plaintext query and datasets.
      If the type is $type\ j$, increase $n_j$ by $1$.
  \item Generate a noise vector $(\hat{n}_0, \hat{n}_1, \dots, \hat{n}_d)$ according to Theorem~\ref{thm:differential_privacy} to achieve differential privacy. The randomness of the noise is sampled from the random tape $R_e$ of the exchanger.
      For each $j \in \{0, 1, \dots, d\}$, increase $n_j$ by $\hat{n}_j$.
  \item Use $rb$ to open $c_b$ and get $x_b$.
  \item For each $D_i$, retrieve $x_{ib}'$ using $gid_b$ and $q$, then check if $c_b = F(x_b, r_b)$ and $x_b = \sum_i x_{ib}'$. Denote the check result by $z$ ($``pass"$ or $\bot$).
\item Output $\vec{v}$ and $z$ to the originator.
\end{enumerate}
}
}
\end{minipage}

We define $\mathcal{F}_{agg}$ as the functionality of secure aggregation as above.
As we stated in Section~\ref{section:anonymous_authorization}, $\Pi_{agg}$ works under the assumption of a semi-honest originator, and the purpose of $\Pi_{auth}$ is to detect a malicious originator that sends unauthorized queries. 
As we can have $\Pi_{agg}$ and $\Pi_{auth}$ share the same PIR query $q$ (this can be guaranteed as the exchanger receives the PIR query in both protocols), we can use $\Pi_{auth}$ to enhance the security of $\Pi_{agg}$. 
We define such a protocol as $\Pi_{auagg}$, which is depicted in Protocol~\ref{protocol:enhanced_aggregation}.
$\Pi_{auagg}$ enables the secure aggregation to work under the assumption of $\mathcal{P}_{\sysname}$. 
In $\Pi_{auagg}$, if the originator's query is authorized, the originator gets the output of $\Pi_{agg}$ and checks the consistency of the commitment from the borrower and the information from the lenders.
On the other hand, when getting a symbol $\bot$ from $\mathcal{F}_{auth}$, the exchanger discards the messages and aborts the protocol, and in this case the originator would get no message except a $\bot$ from the exchanger.

\begin{protocol}[htbp]
    \DontPrintSemicolon
    \begin{enumerate}[label={\alph*)}, leftmargin=0.1cm, rightmargin=0.3cm, itemsep=0mm]
        \item All the participants execute the steps $a$ to $e$ of $\Pi_{agg}$. 
        \item Concurrently, the borrower, the originator and the exchanger call $\mathcal{F}_{auth}$. The query $q$ the originator inputs to $\Pi_{agg}$ is the same as the query used in $\mathcal{F}_{auth}$. The exchanger gets an indicator $z$ which indicates the query is authorized or not.
        \item If $z$ is $``pass"$, the participants execute the remaining steps of $\Pi_{agg}$ and the originator gets the output of $\Pi_{agg}$. Otherwise, the exchanger aborts the execution and the originator gets a symbol $\bot$.
    \end{enumerate}
    \caption{Authorization-enhanced aggregation, $\Pi_{auagg}$}
    \label{protocol:enhanced_aggregation}
\end{protocol}

\begin{thm}
\label{thm:auth_security}
$\Pi_{auagg}$ securely realizes $\mathcal{F}_{agg}$ in the $\mathcal{F}_{auth}$-$hybrid$ model in the presence of $\mathcal{P}_{\sysname}$.
\end{thm}

\noindent \textsc{Proof sketch: }
\para{Correctness. }
If the participants behave honestly, the check of $\Pi_{agg}$ passes due to the bindness of the commitment scheme.
This is because in $\Pi_{agg}$ both $c_b$ and $c$ commit to $\sum_i x_{ib}$, with $r_b$ and $r_z + \sum_i r_i$ as the randomness, respectively. 
As $\Delta r + r_o = r_b - \sum_i r_i - r_z$, the check $c_b = c \cdot h^{\Delta r + r_o}$ indicates the consistency.
The originator's input includes a query $q$ and a group id $gid_b$, and $\mathcal{F}_{auth}$ ensures that an adversary $\mathcal{A}_o$ that corrupts the originator and gives incorrect input to the input tape of the originator would be detected.
Now we analyze the situation where the borrower is corrupted by an adversary $\mathcal{A}_b$ and gives dishonest inputs. As the messages the borrower sends in $\Pi_{agg}$ include a commitment $c_b$ and a random number $\Delta r_b$, $\mathcal{A}_b$ should construct a message pair $(c_b', \Delta r_b')$ satisfing $c_b' = c \cdot h^{\Delta r_b' - r_z + r_o}$, which indicates $c_b' = F(\sum_i x_{ib}, \sum_i r_i) \cdot h^{\Delta r_b' + r_o}$, equivalent to $c_b' = F(\sum_i x_{ib}, \sum_i r_i) \cdot F(0, \Delta r_b' + r_o)$.
Due to the homomorphism property of the commitment scheme, we can see that $c_b'$ is another commitment to $\sum_i x_{ib}$ in this case.
On the other hand, as each $r_i$ is generated using a random seed $\tau_{ib}$,
$F(\sum_i x_{ib}, \sum_i r_i)$ is masked by pseudo-random numbers and a malicious borrower using incorrect seeds can only construct a same commitment with negligible probability.
Therefore, we can conclude that a corrupted borrower using dishonest inputs can only pass the check of the originator in $\Pi_{agg}$ with negligible probability. So we get the correctness.

\para{Privacy. }
For privacy, the exchanger gets an indicator which indicates that the inputs of the originator and borrower are valid or not.
The originator gets an indicator which indicates that the borrower's input is consistent with the lenders' inputs or not. 
In addition, the originator gets a vector $\vec{v}$ which contains the lenders' response types with noise.
The two indicators are necessary for our security goals and do not contain sensitive information.
The type count vector $\vec{v}$ does not reveal the concret values of the borrower's inputs, and protects the types of the lenders' databases with differential privacy.

\para{Security. }
We first consider the semi-honest lenders. The lenders input their databases and seeds, and get no output in $\Pi_{auagg}$.
The view of each lender $S_i$ is a $d$-dimensional PIR query $q$.
We can construct a simulator $\mathcal{S}_i$ that randomly pick a position in the lender's database and generates a valid $d$-dimensional PIR query $q'$ encrypted using a random key.
Given the security of the Paillier crypto system, we have that $q$ and $q'$ are computationally indistinguishable.

The view of the exchanger in $\Pi_{auagg}$ includes a PIR query $q$ from the originator, a commitment $c_b$ and a random number $\Delta r_b$ from the borrower, and a set of PIR responses $C$ from the lenders.
We construct a simulator $\mathcal{S}_e$ that first generates a random commitment $c_b'$, then uses the originator's public key $(n, g)$ to generate a PIR query $q'$, samples a random number $\Delta r_b'$ from the field of the committed values, and generates a set of random PIR responses of $type\ 0$ (denoted by $C'$). Finally, $\mathcal{S}_e$ sends the messages it generates to the exchanger.
We first see that both $c_b$ and $c_b'$ are commitments with independent randomness, and $\Delta r_b$ is masked by pseudo-random numbers generated using a seed unknown by the exchanger, thus is indistinguishable with $\Delta r_b'$ for the exchanger. Further more, as $\Delta r_b$ is masked by $r_o$, the distribution of $\Delta r_b$ is independent of the distribution of $c_b$. Thus we have $(c_b, \Delta r_b) \approx (c_b', \Delta r_b')$.
On the other hand, both $q$ and $q'$ consist of Paillier ciphertexts and are indistinguishable for the exchanger. Meanwhile, the PIR responses in $C$, though generated using $q$, are randomized by the lenders (see Algorithm~\ref{algorithm:pir_sparse_recursive} for details), thus are also indistinguishable with the ciphertexts in $C'$ for the exchanger.
Finally, we have $(c_b, \Delta r_b, q, C) \approx (c_b', \Delta r_b', q', C')$.

For the originator, we assume that it is corrupted by an adversary $\mathcal{A}_o$. 
If $\mathcal{A}_o$ gives incorrect inputs to the originator, $\mathcal{F}_{auth}$ detects it (step 2 in $\Pi_{auagg}$) and the exchanger outputs $\bot$ to the originator and aborts the execution.
In this case $\mathcal{A}_o$ only gets a symbol $\bot$ as output and the view is empty.
But if $\mathcal{A}_o$ gives correct inputs to the originator and still gets $\bot$ from the exchanger, $\mathcal{A}_o$ knows that the borrower lies about her identity, and the view is also empty in this case.
We then consider the case where $\mathcal{A}_o$ gives correct inputs and the exchanger sends the messages from the borrower and the PIR responses to the orignator. 
In this case, the view of $\mathcal{A}_o$ includes a commitment $c_b$, a random number $\Delta r$ and a set of PIR responses $C$ that contains the responses from the lenders and the noise responses from the exchanger. After decrypting the ciphertexts in $C$, $\mathcal{A}_o$ gets the type count vector $\vec{v} = \{n_0, n_1, \dots, n_d\}$ and checks the consistency, the result of which is denoted as $z$. $\vec{v}$ and $z$ are the originator's output.
We construct a probabilistic-polynomial time simulator $\mathcal{S}_{o}$ which works as follows:
\begin{enumerate}[leftmargin=0.5cm, label=\arabic*.]
    \item Receive $\tau_{ob}$, the public key $(n, g)$, $\vec{v}$ and $z$ from $\mathcal{A}_o$.
    \item Sample a random number $r_b'$ from the field of the commited values, and generate a commitment $c_b' = F(0, r_b')$.
    \item Generate fake PIR responses according to $\vec{v}$. Specifically, for $1 \le j \le d$, generate $n_j$ fake responses of $type\ j$, and for $j = 0$, generate $n_0$ responses as follows:
          \begin{enumerate}[label={\alph*)}, leftmargin=0.5cm, rightmargin=0.3cm, itemsep=0mm]
              \item Sample $n_0$ random numbers $r_1, r_2, \dots, r_{n_0}$ from the field of the commited values, and calculate $r_z' = \sum_{i=1}^{n_0} r_{n_0}$.
              \item Generate $n_0$ commitments that commit to $0$ using the random numbers sampled in the previous step. 
              \item Encrypt the commitments as $n_0$ responses of $type\ 0$.
          \end{enumerate}
          Shuffle these fake PIR responses. Denote the set of the responses as $C'$.
    \item If $z$ is $``pass"$, calculate $r_o' = \text{PRF}_{\tau_{ob}}(``rc" \| b \| \text{date})$ and set $\Delta r' = r_b' - r_z' - r_o'$;
          otherwise, sample a random number $r'$ from the field of the commited values and set $\Delta r' = r'$.
    \item Send $c_b'$, $\Delta r'$ and $C'$ to $\mathcal{A}_o$.
\end{enumerate}

\noindent Our goal is to prove that $(c_b, \Delta r, C) \approx (c_b', \Delta r', C')$.
We first have $\Delta r \approx \Delta r'$, as $\Delta r'$ is a random number generated by $\mathcal{S}_o$, while $\Delta r$ is masked by a random number generated by the borrower.
On the other hand, $\mathcal{A}_o$ gets two things from $C$ (\emph{resp.} $C'$): a type count vector $\vec{v}$ (\emph{resp.} $\vec{v}'$) and a list of commitments $c_1, c_2, \dots, c_{n_0}$ (\emph{resp.} $c_1', c_2', \dots, c_{n_0}'$).
We can see that the consistency check result in the simulation (denoted as $z'$) equals $z$.
This is because $\mathcal{A}_o$ checks the consistency by comparing $c_b'$ with $\prod_{j=1}^{n_0} c_j' \cdot h^{\Delta r' + r_o'}$.
When $z$ is $``pass"$, $r_o'$ is a pseudo-random number calculated using $\tau_{ob}$: $r_o' = \text{PRF}_{\tau_{ob}}(``rc" \| b \| \text{date})$. In this case, $\Delta r' + r_o' = r_b' - r_z'$ and the consistency check passes, indicating that $z' = ``pass"$.
When $z$ is $\bot$, $\Delta r'$ is a random number and the consistency check fails, indicating that $z' = \bot$.
In both cases, we can threat the tuple $(c_b, \Delta r, C)$ (\emph{resp.} $(c_b', \Delta r', C')$) as the output of a randomized function $f$ which takes $C$ (\emph{resp.} $C'$) as input.
Thus if we can prove $C \approx C'$, we can get $(c_b, \Delta r, C) \approx (c_b', \Delta r', C')$.   

It remains to show $C \approx C'$. First, as $\mathcal{S}_{o}$ generates responses according to $\vec{v}$, we have $\vec{v} = \vec{v'}$. 
Then we argue that $(c_1, c_2, \dots, c_{n_0}) \approx (c_1', c_2', \dots, c_{n_0}')$. 
Actually, as these commitments are generated independently using truely random numbers or pseudo-random numbers with seeds unknown by $\mathcal{A}_o$, the commitments are indistinguishable for $\mathcal{A}_o$. 
Thus we have $C \approx C'$ for $\mathcal{A}_o$, which implies that $(c_b, \Delta r, C) \approx (c_b', \Delta r', C')$.

Finally, we take the borrower into account. As the borrower receives no messages in $\Pi_{auagg}$ when treating anonymous authorization as a functionality, the security for the borrower is trival.
We then consider an adversary $\mathcal{A}_{ob}$ that corrupts both the originator and the borrower.
With the inputs from the borrower, $\mathcal{A}_{ob}$ can open the commitments from the lenders.
If the borrower has borrowed money from a lender $S_i$, $\mathcal{A}_{ob}$ surely knows that the type of $S_i$'s PIR response is $type\ 0$.
But if the borrower has not borrowed money from $S_i$, we should prove that our potocol protects the exact type of the PIR response from $S_i$.
Actually, as the response is randomized in Algorithm~\ref{algorithm:pir_sparse_recursive}, $\mathcal{A}_{ob}$ cannot distinguish the ciphertext of the response from the ciphertexts generated by the exchanger. Thus the exact type is still perturbed by the noise with differential privacy.
$\qed$

\subsection{Security Proof of \sysname}

\noindent \begin{minipage}[h]{0.47\textwidth}
\fbox{
\parbox{\textwidth}{
\begin{center}
Functionality $\mathcal{F}_{\sysname}$
\end{center}

\para{Inputs: }
The borrower inputs a number $x_b$. 
The originator inputs a plaintext query $q$, a function $f$ and a number $t$.
Each lender $S_i$ inputs its database $D_i$.

\para{Process: }
\begin{enumerate}[leftmargin=0.5cm, label=\arabic*.]
    \item For each lender $S_i$, select a number $x_i$ using $q$. Calculate $x = \sum_i x_i$.
    \item Check that $x_b$ equals $x$. If the check fails, output a symbol $\bot$ to the originator and abort.
    \item Calculate $y = f(x_b; t)$ and output $y$ to the originator.
\end{enumerate}
}
}
\end{minipage}

\begin{protocol}[htbp]
    \DontPrintSemicolon
    \begin{enumerate}[label={\alph*)}, leftmargin=0.1cm, rightmargin=0.3cm, itemsep=0mm]
        \item The borrower and the originator call $\mathcal{F}_{f}$: the borrower inputs a random number $r_b$ and a commitment $c_b$, while the originator inputs a function $f$ and a number $t$.
        \item Concurrently, all the participants  call $\mathcal{F}_{agg}$. The commitment $c_b$ the borrower inputs to $\mathcal{F}_{agg}$ is the same as the commitment sent to $\mathcal{F}_{f}$. The originator gets an indicator $z$ and a type count vector $\vec{v}$ from $\mathcal{F}_{agg}$.
        \item If $z$ is $``pass"$, the originator uses the output of $\mathcal{F}_{f}$ as the result. Otherwise, the originator aborts the execution and outputs a symbol $\bot$.
    \end{enumerate}
    \caption{\sysname protocol, $\Pi_{\sysname}$}
    \label{protocol:final}
\end{protocol}

\newpage
We finally define the functionality of \sysname as $\mathcal{F}_{\sysname}$, and $\Pi_{\sysname}$ depicts our final composed protocol.
We also use $\mathcal{F}_{f}$ to represent the functionality of the specified evaluation function $f$, which is one of the functions for secure evaluation described in Section~\ref{section:secure_evaluation}.
As the protocols for these functions are existing approaches, we omit the analysis for them in this paper, and focus on the composition of the above functionalities.
The common trait of the functions in Section~\ref{section:secure_evaluation} is that each of them uses a commitment which commits to $x_b$ from the borrower as input.
In $\Pi_{\sysname}$, $\mathcal{F}_{f}$ and $\mathcal{F}_{agg}$ share the same commitment $c_b$ (this can be guaranteed in our protocol as the originator receives $c_b$ in both $\Pi_{auagg}$ and the realization of $f$). Then we have the following theorem:

\begin{thm}
\label{thm:final_security}
$\Pi_{\sysname}$ securely realizes $\mathcal{F}_{\sysname}$ in the $\mathcal{F}_{agg}$-$hybrid$ model in the presence of $\mathcal{P}_{\sysname}$.
\end{thm}

\noindent \textsc{Proof sketch: }
\para{Correctness. }
If the participants behave honestly, $\mathcal{F}_{agg}$ outputs $``pass"$ to the originator and $\Pi_{\sysname}$ outputs the result of $f$ to the originator.
On the other hand, $\mathcal{F}_{agg}$ and $\mathcal{F}_{f}$ ensure that dishonest inputs from the borrower or the originator would be detected without revealing sensitive information from others.

\para{Privacy. }
The output of $\Pi_{\sysname}$ includes the output of $\mathcal{F}_{agg}$ and $\mathcal{F}_{f}$.
We have analyzed the privacy of the output of $\mathcal{F}_{agg}$, while the output of $\mathcal{F}_{f}$ is exactly the originator wants to get from the protocol, and no more information is revealed, which gives the privacy.

\para{Security. }
As $\Pi_{\sysname}$ calls $\mathcal{F}_{agg}$ and $\mathcal{F}_{f}$, the borrower and the lenders gets no output, while the exchanger and the originator get outputs from $\mathcal{F}_{agg}$ and $\mathcal{F}_{f}$.
There is no intermediate messages, and the construction of simulators is trival.
$\qed$


\end{document}